\documentclass[longbibliography,showpacs,floatfix,superscriptaddress,twocolumn,aps,prl]{revtex4-2}
\usepackage[figuresright]{rotating}  
\usepackage{amssymb}
\usepackage{bm}
\usepackage{amsmath}
\usepackage{mathtools}
\usepackage{psfrag}
\usepackage{multirow}
\usepackage{tabularx}
\usepackage{textcomp}
\usepackage{units}
\usepackage{lipsum}
\usepackage{wasysym} 
\usepackage{graphicx}

\usepackage{physics}
\usepackage{soul}
\usepackage{titlesec}
\usepackage{times}
\usepackage[colorlinks=true,citecolor=blue,linkcolor=magenta,hypertexnames=false]{hyperref}
\usepackage{enumitem}
\usepackage{floatrow}

\usepackage[caption=false]{subfig}

\usepackage{tikz}

\renewcommand{\selectlanguage}[1]{}

\begin{document}

\title{Irrational moments and signatures of higher-rank gauge theories in diluted classical spin liquids}
\author{R. Flores-Calderón}
\affiliation{Max Planck Institute for the Physics of Complex Systems, Nöthnitzer Strasse 38, 01187 Dresden, Germany}
\affiliation{Max Planck Institute for Chemical Physics of Solids, Nöthnitzer Strasse 40, 01187 Dresden, Germany}
\author{Owen Benton}
\affiliation{Max Planck Institute for the Physics of Complex Systems, Nöthnitzer Strasse 38, 01187 Dresden, Germany}
\affiliation{School of Physical and Chemical Sciences, Queen Mary University of London, London, E1 4NS, United Kingdom}
\author{Roderich Moessner}
\affiliation{Max Planck Institute for the Physics of Complex Systems, Nöthnitzer Strasse 38, 01187 Dresden, Germany}

\begin{abstract}
Classical spin liquids (CSLs) have proved to be a  fruitful setting for the emergence of exotic  gauge theories. Vacancy clusters in CSLs can introduce gauge charges into the system, and the resulting behavior in turn reveals the nature of the underlying theory. We study these effects for a series of CSLs on the honeycomb lattice. We find that dilution leads to the emergence of effective free spins with tuneable, and generally irrational, size. For a specific higher-rank CSL, described by a symmetric tensor gauge fields, dilution produces {\it non-decaying} spin textures with a characteristic quadrupolar angular structure, and infinite-ranged interactions between dilution clusters.
 \end{abstract}
\maketitle

\textit{Introduction}.-- Strongly interacting phases of matter allow the investigation of exotic field theories apparently not present at the fundamental level in our universe. Spin liquids, in particular, are known as phases of magnetic matter realizing emergent gauge theories with fractionalized excitations \cite{anderson_resonating_1973,balents_spin_2010,zhou_quantum_2017,wen_quantum_2002,castelnovo_spin_2012,FieldGuide}. 
Classical spin liquids (CSLs) are found in classical spin models with highly degenerate ground states, subject to local constraints \cite{moessner_low-temperature_1998, moessner_pyrochlore_1998}.
This constraint can often be understood as an effective Gauss law
\cite{isakov_dipolar_2004,henley_power-law_2005,yan_classification1_2023, yan_classification2_2023},
of an emergent gauge theory.

The set of gauge theories realized in spin liquids  now  extends to higher-rank cases, where gauge fields are  symmetric tensors of rank $>1$ \cite{xu_gapless_2006,chamon_quantum_2005,haah_local_2011,vijay_fracton_2016,vijay_new_2015,halasz_fracton_2017,you_building_2019,xu_bond_2007,xu_emergent_2010}. 
Excitations of these spin liquids often have restricted mobility as a result of the conservation laws of the gauge theory \cite{pretko_generalized_2017, pretko_subdimensional_2017};
immobile quasiparticles are known as fractons \cite{Pretko_2020,nandkishore_review}. They have been studied extensively due to their connections with fault tolerant quantum computing \cite{kubica_ungauging_2018,schmitz_recoverable_2018}, the theories of elasticity \cite{pretko_fracton-elasticity_2018}, and gravity \cite{xu_gapless_2006,pretko_generalized_2017,benton_spin-liquid_2016}, exotic topological orders \cite{vijay_fracton_2016,xu_bond_2007,xu_emergent_2010} and  holography \cite{yan_hyperbolic_2019}. 
Multiple instances of CSL models exhibiting higher rank gauge fields are now known \cite{yan_rank--2_2020, benton_topological_2021, yan_classification1_2023}.

CSLs exhibit an analogue of fractionalization, revealed through their response to dilution \cite{moessner_magnetic_1999, henley_effective_2001, sen_fractional_2011, sen_vacancy-induced_2012,laforge_quasispin_2013, sen_topological_2015}.
When a cluster of vacancies is introduced such that one of the local constraints of the clean spin liquid has only one remaining spin, this spin becomes an ``orphan'', which responds to external fields like a free spin with a fractional moment.
If the local constraint can be understood as a Gauss
law, the orphan spins represent a localized gauge charge for which the energy cost of creation has already been paid by the quenched dilution itself. 
These induced charges  interact with one another,  mediated by the correlations of the spin liquid.

Fractional orphan spin moments with  half of the bare moment are known to arise in frustrated models on lattices of corner-sharing simplices \cite{moessner_magnetic_1999,henley_effective_2001,sen_vacancy-induced_2012,sen_topological_2015, Patil2020}. In contrast, the maximally frustrated honeycomb lattice exhibits a fractional orphan spin moment with a size one third of the bare moment \cite{rehn_classical_2016}.
These fractional moments have been invoked \cite{sen_fractional_2011} to understand the observation of a Curie tail in the susceptibility on the frustrated magnetic material SrCr$_{9p}$Ga$_{12-p}$O$_{19}$ \cite{schiffer_two-population_1997, limot02}, where the term orphan spin was coined \cite{schiffer_two-population_1997}.
How  orphan spin behaviour generalizes to more complex spin liquids, including those with higher rank gauge fields, has not been addressed.

 \begin{figure}[ht!]
    \centering\includegraphics[width= 0.9\columnwidth]{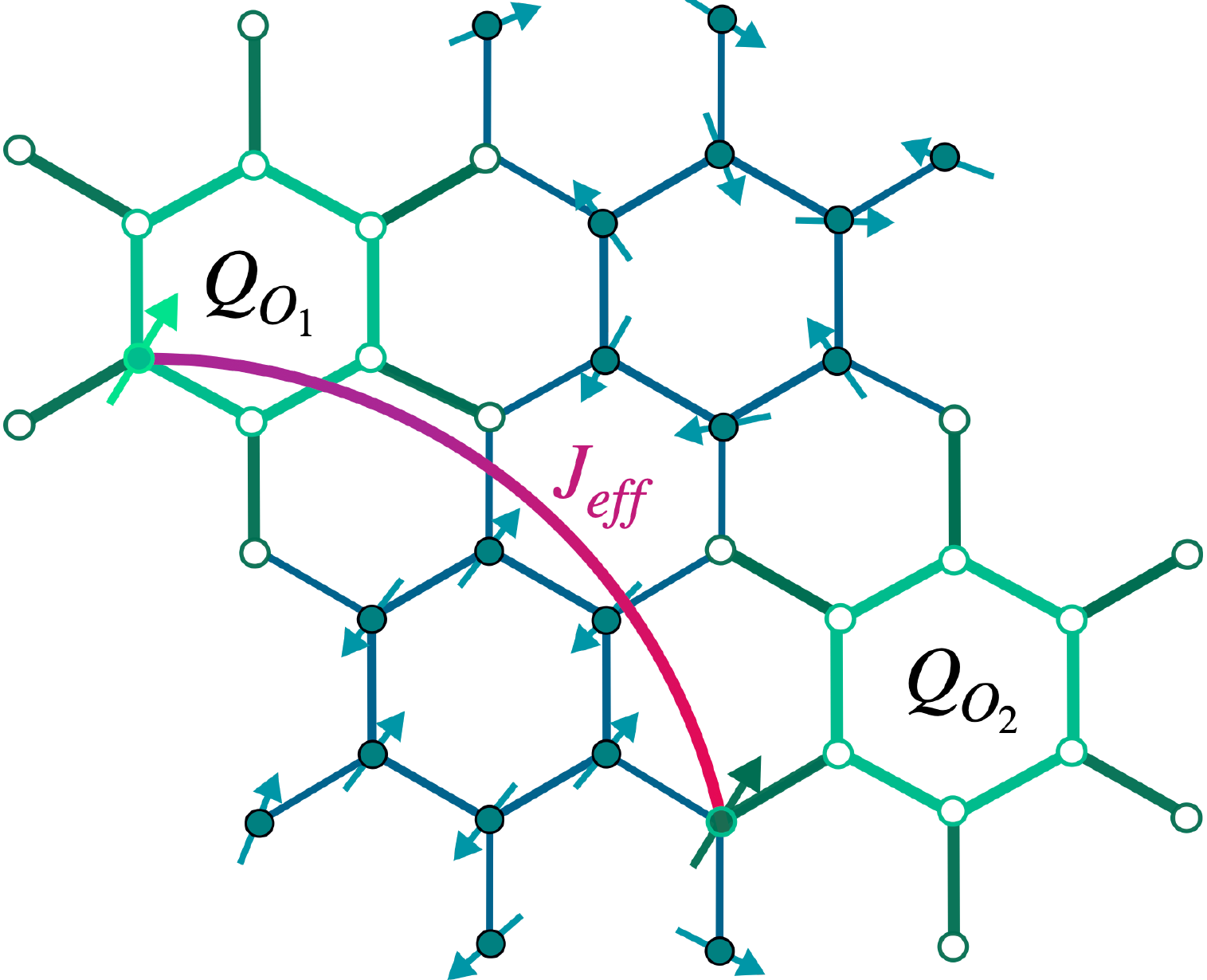}
    \caption{
    Interactions and orphan spins in the honeycomb-snowflake model \cite{benton_topological_2021, yan_classification1_2023}.
    The Hamiltonian [Eq. (\ref{SnowHoneyHam})] enforces a local constraint in which the sum of spins around each hexagon (light green), added to $\gamma$ times the sum of spins connected to the exterior of the hexagon (dark green) must vanish.
    The snowflakes labelled $Q_{O_1}$ and $Q_{O_2}$ show two distinct vacancy
    configurations, with one  orphan spin in the interior of the orphan snowflake (bonds shown in light green), with the other on its boundary (bonds shown in dark green).}
    \label{fig:Schem}
 \end{figure}

In this Letter we study a generalized model for CSLs on the honeycomb lattice \cite{benton_topological_2021},  finding a {\it continuously tunable} orphan spin fractionalization, which therefore allows for the generic appearance of irrational moments. This tunability occurs as a consequence of a modified ground state constraint which weighs different spins on a cluster differently. 
This behaviour is not restricted to 
the particular model discussed here:
the same mechanism will also be operative in simpler CSLs with variable couplings enforcing the ground state constraints, such as on breathing lattices \cite{inprep}.

Having analysed the general case, we   
specialize to a particular point of the phase diagram \cite{benton_topological_2021}, at which a higher-rank spin liquid emerges. We study the introduction of orphan spins at this point, as a means to access the response of the spin liquid to forcing fractonic charges into the system. We find that the  spin texture surrounding the orphan spin snowflake has a characteristic quadrupolar angular dependence, and does not decay at large distances. Furthermore the effective  interaction between orphan spin snowflakes is thermally screened on a scale $\sim T^{-\frac{1}{4}}$, as opposed to $T^{-\frac{1}{2}}$  for conventional orphan spins \cite{sen_fractional_2011}. We thus  determine   features of the orphan spin behaviour characteristic of emergent higher-rank gauge theories.

\floatsetup[figure]{style=plain,subcapbesideposition=top}

\textit{Model}.-- We consider the ``honeycomb-snowflake" model of classical $O(3)$ spins $\vec{S}_i$ on the honeycomb lattice \cite{benton_topological_2021, yan_classification1_2023} 
\begin{align}
    \mathcal{H}=\frac{J}{2} \sum_{\varhexagon} \left(\sum_{i \in \varhexagon} \vec{S}_i+\gamma \sum_{i \in \langle\varhexagon\rangle} \vec{S}_i \right)^2-\sum_{i} \vec{S}_{i}\cdot\vec{h}, \label{SnowHoneyHam}
\end{align}
with antiferromagnetic $J>0$, $\gamma$ a dimensionless, tunable, parameter, and  magnetic field $\vec{h}$.
In zero field, the sum inside parentheses defines the ground state constraints, with the first a sum over spins on each hexagonal plaquette, and the second over spins adjacent to the plaquette (see Fig.~\ref{fig:Schem}); together, these make up the snowflake.
Any configuration where this sum vanishes for all hexagons of the lattice is a ground state. This yields a correlated ground state manifold of extensive dimensionality, around which the system fluctuates at low temperature.

Varying $\gamma$ tunes the ground state constraint resulting in a rich phase diagram \cite{benton_topological_2021}.  
$\gamma=0$ corresponds to a model of maximally frustrated hexagonal plaquettes, studied in \cite{rehn_classical_2016}. 
$\gamma=1/2$ realises a rank-2 $U(1)$ spin liquid, with a low energy theory in terms of a traceless symmetric tensor field. 

The continuous interpolation between multiple phases allows us to study within one model the generic response of spin liquids to dilution. 
After discussing the general effect of dilution in the model as a function of $\gamma$, we will focus on particular signatures linked to thie higher rank spin liquid at $\gamma=1/2$.
 
\textit{Continuously tunable orphan moment}.-- We consider vacancies placed at fixed positions of the honeycomb lattice. A constrained snowflake with only a few  spins removed is still able to fulfil the ground state constraint $\vec{S}_\gamma=0$. 
However, if the vacancies are placed in such a way that only a single (orphan) spin remains in a given snowflake, the sum will always give a nonzero vector, of magnitude $S$ ($\gamma S$) if the orphan spin is located on the hexagon (on the boundary of the snowflake), as shown schematically in Fig. \ref{fig:Schem}. The failure to satisfy the constraint allows us to identify the presence of an orphan spin with the presence of a gauge charge. Its presence  affects the response to external fields. In other CSL models, it has been observed that orphan spins respond like  free spins with the surrounding spin liquid renormalizing the magnetic moment to be fractional: $S/2$ for corner sharing lattices and $S/3$ for the Honeycomb model with $\gamma=0$ \cite{sen_vacancy-induced_2012,rehn_classical_2016}. We first analyze the fate of this fractionalization as $\gamma$ is varied.

We do this by means of the vacancy field theory, developed in \cite{sen_vacancy-induced_2012}, which we also compare with the results of classical Monte Carlo simulations.
The vacancy field theory relies on the self-consistent Gaussian approximation (SCGA) which can be viewed as the leading order of a large-$\mathcal{N}$ expansion, where $\mathcal{N}$ is the number of spin components.
This method treats the spin normalization constraints on average ($\langle \vec{S}_i \cdot \vec{S}_i \rangle=S^2$) for all spins outside the orphan cluster, while fixing the spin at
the vacancy sites to be exactly zero (not only on average), and also treating the normalization of the orphan spin exactly.
We calculate the magnetization as a function of $h$ and extract the orphan spin contribution by subtracting the result without dilution $M_{ud}$ from the result in the presence of dilution $M_d$.

The details of the field theory calculation are given in the Supplemental Material \cite{supplemental}. For an orphan spin on the interior of a hexagon, the orphan spin magnetisation is found to follow the functional form expected for a free spin of length $\alpha S$
\begin{equation}
M_d - M_{ud} = \alpha S \left( \coth(\alpha S \beta h) - \frac{1}{\alpha S \beta h} \right) = \alpha S L(\alpha S)
\label{eq:langevin}
\end{equation}
with $\beta$ being the inverse
temperature and 
\begin{equation}
1/\alpha={3(1+\gamma)}
\label{eq:fraction}
\end{equation}
This reproduces the known case of $\alpha(\gamma=0)=1/3$~\cite{rehn_classical_2016}.
If the orphan spin is located on the boundary of the snowflake, the same result holds, with $\alpha\rightarrow\alpha\gamma$.

\begin{figure}
    \centering
     \sidesubfloat[]{\includegraphics[width= 0.95\columnwidth]{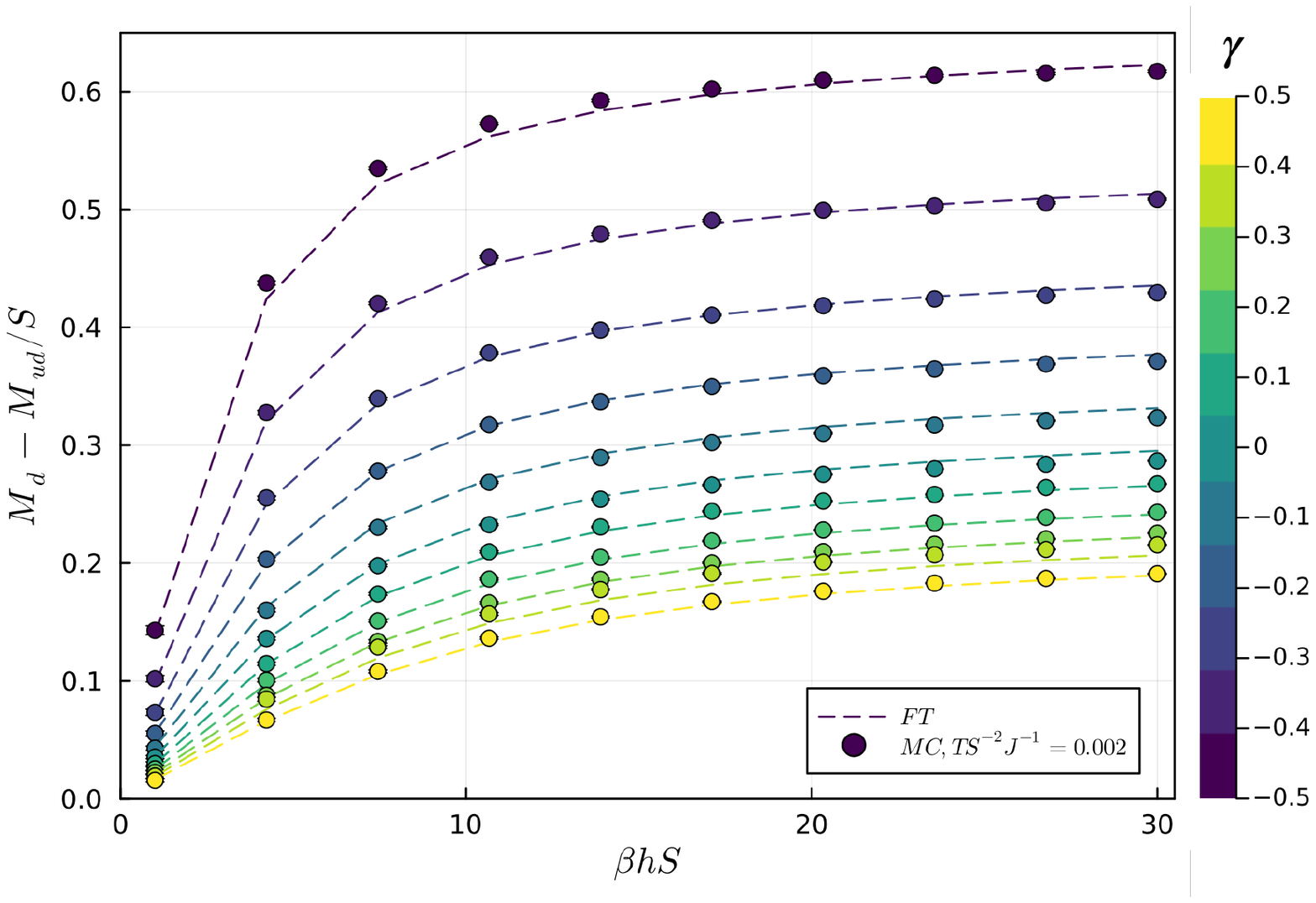}\label{fig:Mag_a}}\\
     \sidesubfloat[]{\includegraphics[width=0.95\columnwidth]{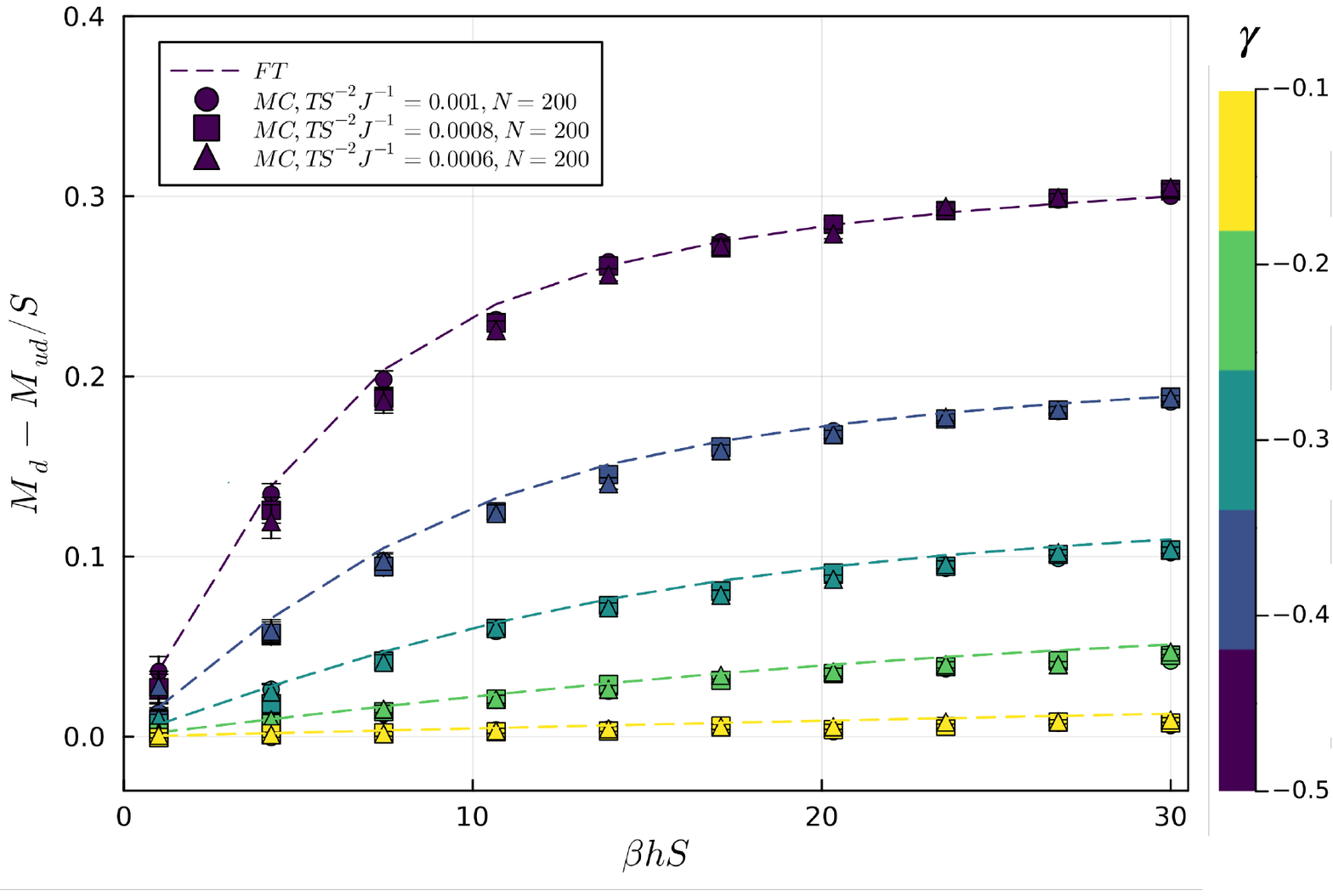}\label{fig:Mag_b}}
    \caption{Orphan spin magnetisation, comparing field theory calculations (dashed lines) with the results of Monte Carlo simulations (points).
    The orphan spin magnetisation is extracted by taking the difference
     between the diluted $M_{d}$ and undiluted $M_{ud}$ magnetization of the honeycomb-snowflake model as a function of magnetic field $h$ and inverse temperature $\beta$. Results are shown for several values $\gamma$, indicated by the color scale. Panel (a) shows the results for an orphan spin located on the interior of an orphan snowflake ($Q_{O_1}$ in Fig. \ref{fig:Schem}), with panel (b) showing the result for an exterior orphan ($Q_{O_2}$). There is good agreement between theory and simulation for both types of orphan and values of $\gamma$. Data is shown for multiple temperatures in (b), to establish that the data collapses when plotted as a function of $\beta h$, as expected for an effectively free spin.} 
    \label{fig:Mag}
\end{figure}

To verify this result, we have performed Monte Carlo simulations of the model Eq.~\ref{SnowHoneyHam}, using the heat bath algorithm.
As expected from Eq. \ref{eq:langevin}, the simulated magnetisation from different temperatures collapses when plotted as a function of $\beta h$, as shown in Fig. \ref{fig:Mag_b}. Comparison of  simulation with field theory  produces good agreement, Fig. \ref{fig:Mag}. We note that Eq. (\ref{eq:fraction}) implies that $\alpha$ can be varied continuously by varying $\gamma$, and is thus not restricted to the rational values found in previous works \cite{sen_fractional_2011, sen_vacancy-induced_2012, rehn_classical_2016, rehn_fractionalized_2017}.
The honeycomb CSL model of Eq. (\ref{SnowHoneyHam}) thus establishes irrational orphan spin fractionalization.

\begin{figure}
    \centering
     \sidesubfloat[]{\includegraphics[width=0.5
     \columnwidth]{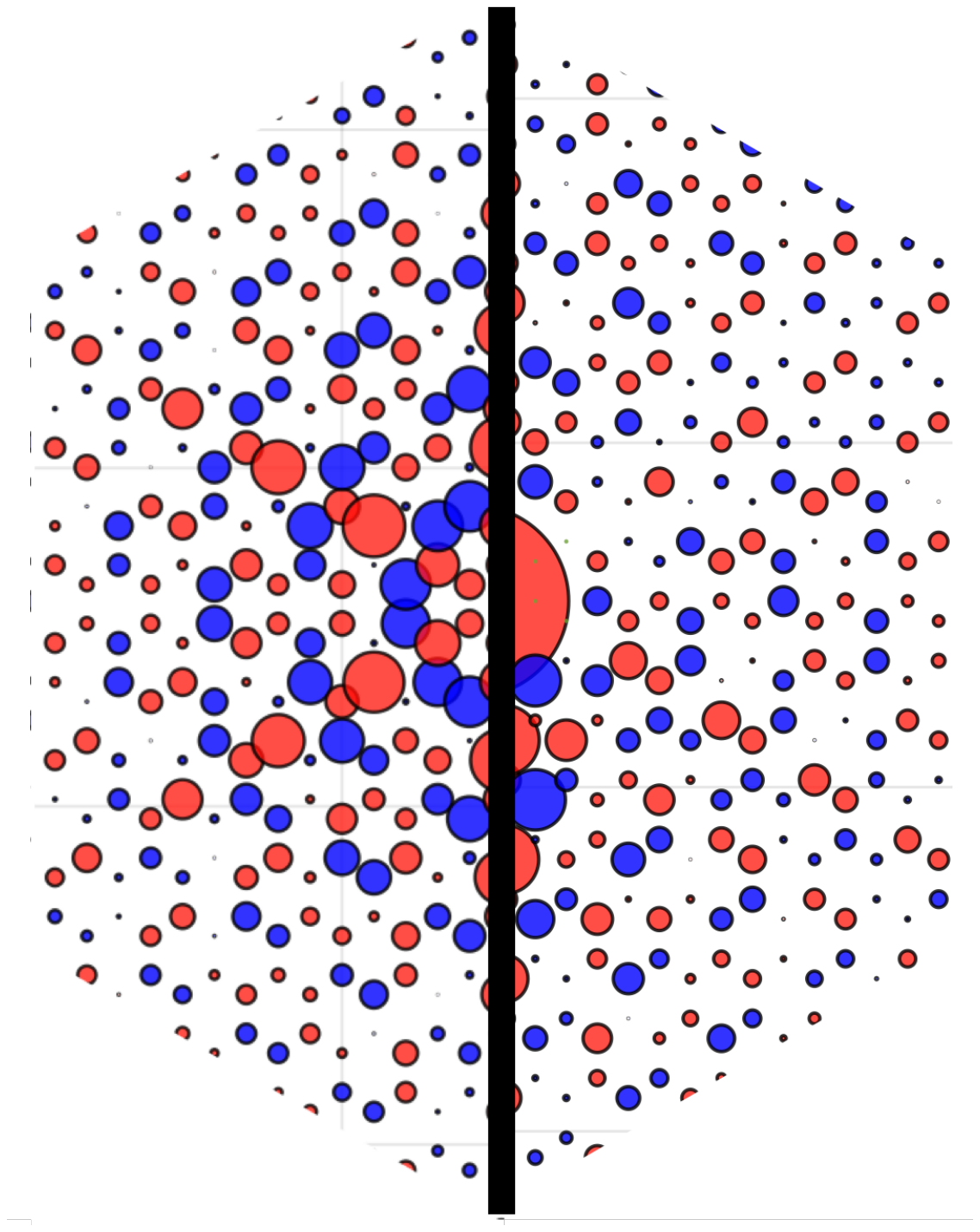}}\\
     \sidesubfloat[]{\includegraphics[width=0.9\columnwidth]{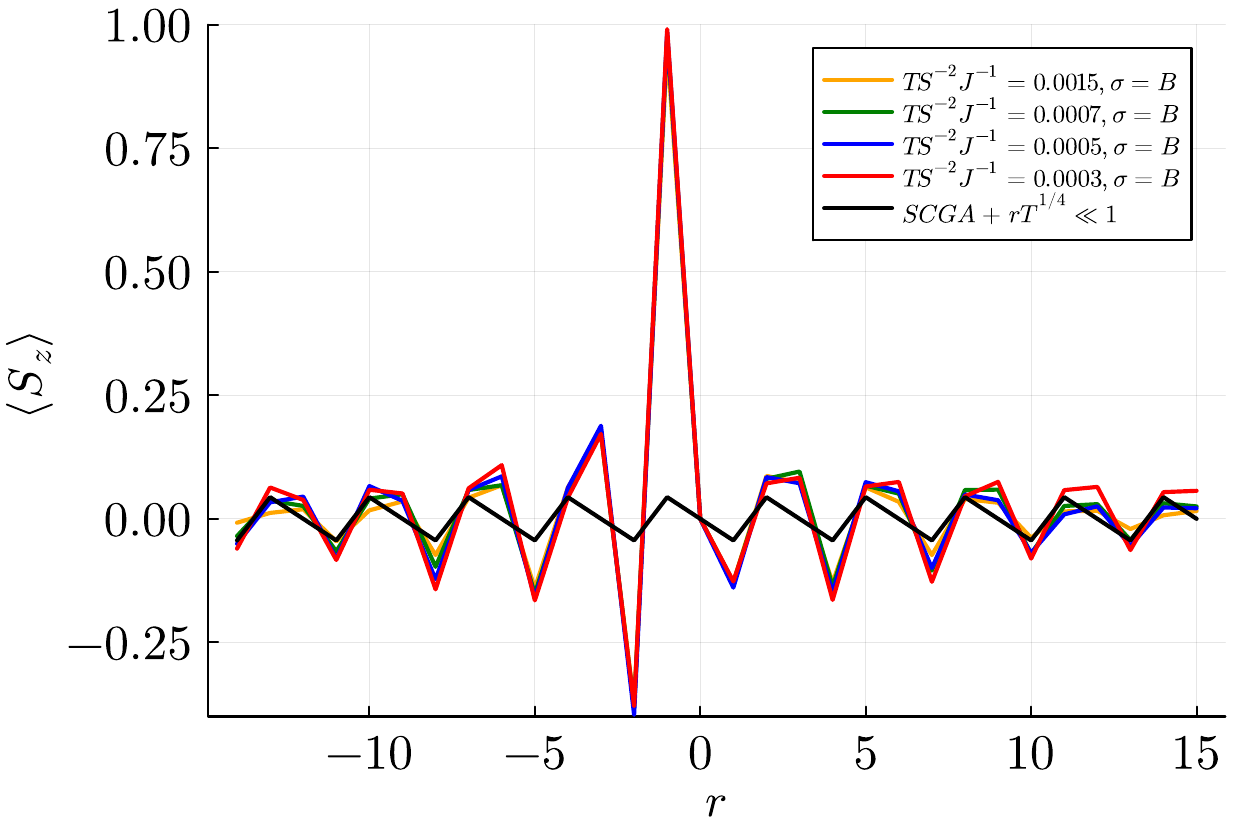}}
    \caption{Spin texture around an orphan spin $O_1$ (Fig. \ref{fig:Schem}) as measured by the expectation value $\expval{S^z(\vec{r})}$, for $\gamma=1/2$ in the honeycomb snowflake model.
     a) Monte Carlo simulations (right) and  far-field $(\abs{\vec{r}-\vec{r}_O}\gg a)$  result from the vacancy field theory (left) as described in the Supplementary Material. The radius of the circle is proportional to the magnitude, while the color indicates the sign (blue $>0$, red $<0$). A strongly angular dependent texture is observed up to a thermal length scale $\sim T^{-1/4}$. Within this scale, the texture is non-decaying. The differences are due to the internal structure of the orphan snowflake, which is neglected in the field theory calculation. b) Cut along the x direction for the $B$ sub-lattice showing agreement of Monte Carlo simulations with the   field theory result at long distances.}
    \label{fig:SpinText}
\end{figure}

\textit{Higher-rank gauge theory}.-- As the orphan magnetic moment varies continuously,  it is worth studying qualitative signatures in the orphan physics to reveal more clearly the nature of the low energy gauge theory describing the spin liquid. In particular we focus on $\gamma=1/2$, where a rank-2 $U(1)$ spin liquid described by a traceless symmetric tensor $m_{\mu\nu}$ has been predicted \cite{benton_topological_2021}. In this case the orphan carries a gauge charge, namely a source of the higher rank divergence $\partial_\mu \partial_\nu m_{\mu\nu}\neq 0$. We consider the spin texture generated by this gauge charge, shown in Fig. \ref{fig:SpinText}. We present both the Monte Carlo as well as the field theory results, with the latter only valid in the far field limit, $\abs{\vec{r}-\vec{r}_O}\gg a$ for $a$ lattice spacing, $\vec{r}$ the measurement position and $\vec{r}_O$ the orphan position. The $z$ component of the spin texture displays a symmetric pattern localized near the orphan spin, with a strong sub-lattice dependence. The field theory calculation, see Supplementary Material \cite{supplemental}, relates the  texture to the correlator between the spin and  charge as
 \begin{align}
     \expval{S^z(\vec{r})}=\dfrac{1}{2}\alpha^2 h +\dfrac{1}{3}\beta \expval{\vec{S}_\gamma(\vec{r}_O)\cdot \vec{S}(\vec{r})}_{ud} \ L \left( \alpha \beta h \right).
 \end{align}
The correlator is calculated from the undiluted theory and $L(x)$ is the 
Langevin function given in Eq. (\ref{eq:langevin}).

The decay of the spin texture with distance is controlled by temperature, with thermally excited violations of the spin liquid constraints exponentially screening the gauge charge on a  lengthscale $\xi(T)$.
In contrast to conventional CSLs with $\xi(T) \sim T^{-1/2}$,  the higher rank theory yields $\xi(T)\propto T^{-1/4}$:
\begin{align}
    \expval{\vec{S}_\gamma(\vec{r}_1)\cdot \vec{S}(\vec{r}_2)}_{ud} \propto T F_1((\vec{r}_{1}-\vec{r}_{2}) T^{1/4})\ ,
\end{align}
where $F_1(\vec{x})$ is smooth and decouples into radial and angular parts. For $a\ll r\ll\xi(T)$ the texture does not decay with distance but oscillates with azimuthal angle $\theta$:
\begin{align}
    F_1(\vec{r}) \propto \cos{(\vec{K}\cdot \vec{r}+2\eta(\vec{r}) \theta)},
    \label{eq:nondecay}
\end{align}
here $\vec{K}$ is the momentum of the $K$ point in the honeycomb Brillouin zone, as the tensor fields encode antiferromagnetic fluctuations of the spins near the $K$ point of reciprocal space \cite{benton_topological_2021}, $\eta(\vec{r})=\pm 1$ encodes which of the two honeycomb sublattices a spin sits on. 

As $T\to0$, the angular dependence persists at large
distances from the diluted snowflake, Eq. (\ref{eq:nondecay}),  confirmed by simulations, Fig. \ref{fig:SpinText}. The angular dependence arises from  the higher order derivative form of Gauss law--indeed, the $2 \theta$ modulation is just the angular part of the electric field of a quadrupole in usual electromagnetism. The non-decaying behaviour can be linked to the fact that in two dimensions a charge of the higher-rank gauge theory satisfies effectively $\partial_{\mu}\partial_\nu m_{\mu \nu}=q\delta^2(\vec{r})$, with the magnitude of the charge $q$ fixed by microscopics. Simple dimensional analysis then implies absence of a length dependence, i.e.\ a non-decaying solution, in $m$.

We now proceed to study the effective interaction between two orphan spins, shown schematically in Fig. \ref{fig:Schem}. The orphans, both placed in the inner hexagon,  are separated by $\vec{r}_{12}=\vec{r}_1-\vec{r}_2$. From the field theory we see that in the  field limit, which neglects the internal structure of the spin, this reduces to calculating the charge-charge correlator of the undiluted spin liquid:

\begin{align}
      \expval{\vec{S}_\gamma(\vec{r}_1)\cdot \vec{S}_\gamma(\vec{r}_2)}_{ud}  \propto T^{3/2} F_2(\abs{\vec{r}_{1}-\vec{r}_{2}} T^{1/4})\ .
\end{align}
again exhibiting a changed thermal length $T^{-1/4}$ compared to the rank-1 $U(1)$ theory \cite{sen_vacancy-induced_2012}. 
For distances $a\ll {r}_{12}\ll\xi(T)$, the effective
interaction having integrated out the spin liquid is:

\begin{align}
    \beta J_{\text{eff}}(\vec{r}_{12}) \approx  \frac{-\beta^2}{3}\expval{\vec{S}_\gamma(\vec{r}_1)\cdot \vec{S}_\gamma(\vec{r}_2)}_{ud}\propto T^{-\frac{1}{2} }\cos( \vec{K}\cdot \vec{r}_{12} ).
\end{align}

The only  angular dependence comes now from the lattice structure of the theory. The lack of quadrupolar angular dependence in $J_{\text{eff}}$, despite its presence in the spin texture, can be understood by viewing the spin texture as analagous to a wave function, while the effective interaction maps to an overlap of two such wave functions~\cite{rehn_random_2015}, whose angular parts cancel in the overlap.
The same intuition (and calculation) implies the absence of decay of $J_{\text{eff}}$ with distance. 

 \begin{figure}
    \centering
     \sidesubfloat[]{\includegraphics[width=1\columnwidth]{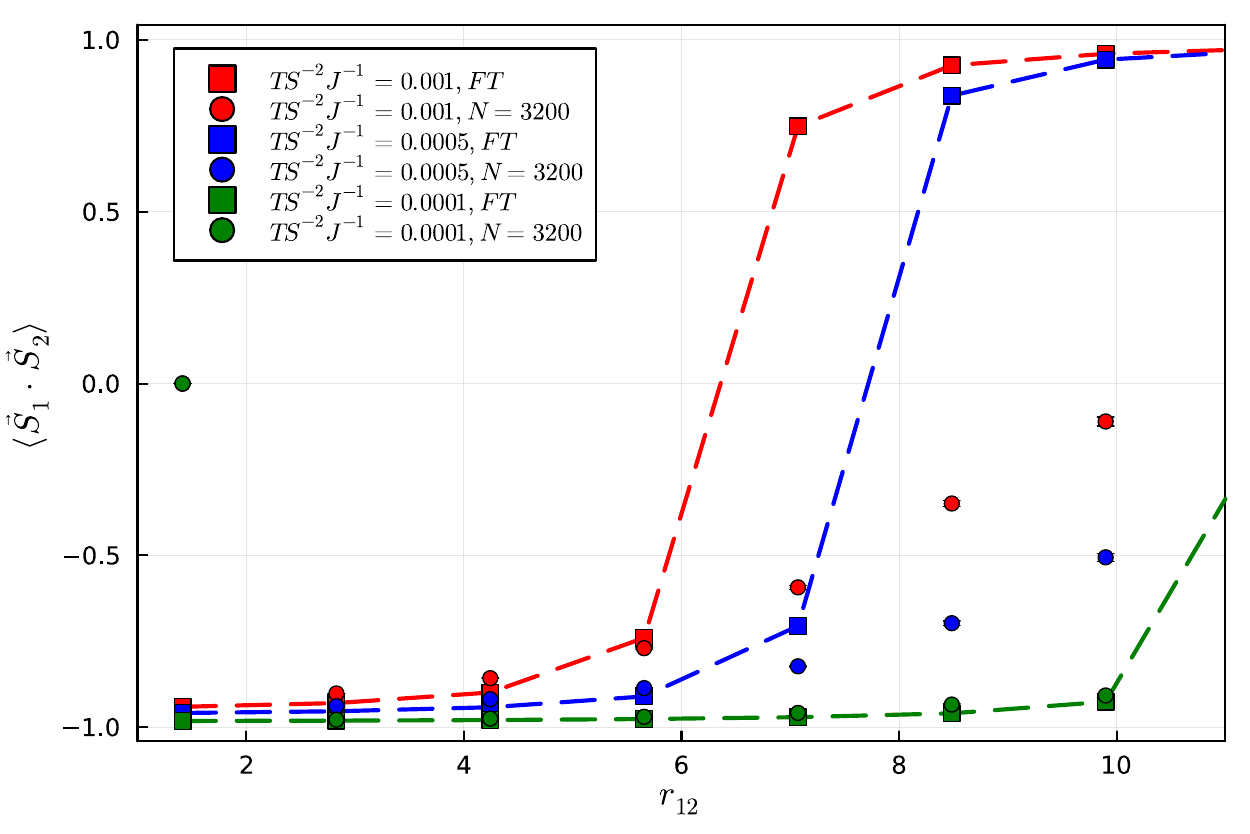}}\\
     \sidesubfloat[]{\includegraphics[width=1\columnwidth]{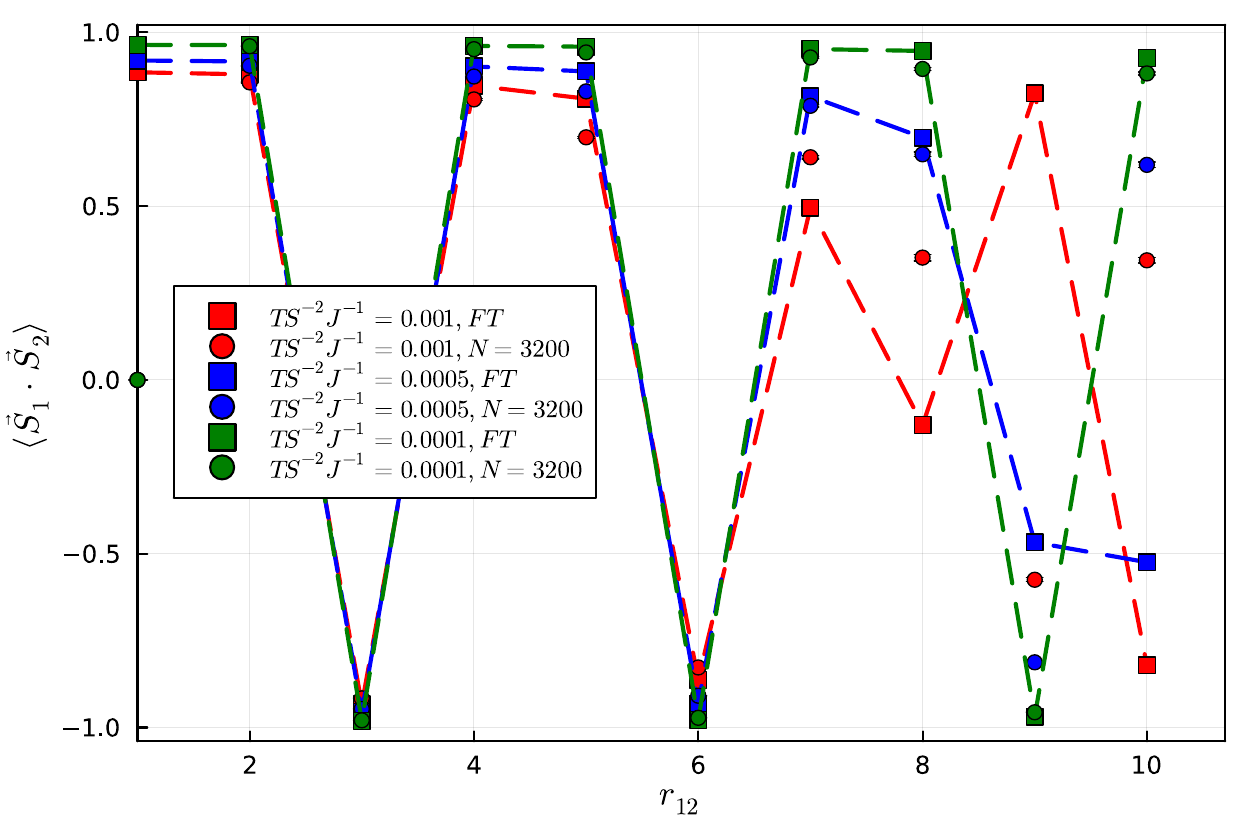}}
    \caption{
    Spin correlations between a pair of orphan spins in the honeycomb-snowflake model for $\gamma=1/2$.
    Correlations are evaluated for vacancy configurations with orphan spins located at varying separations along the (11) (panel (a)) and (10) (panel (b)) directions of the honeycomb lattice. The squares show the hybrid long-wavelength field theory (FT) result while the circles show the Monte Carlo result, different colors indicate different temperatures. Monte Carlo and field theory results agree for distances up to a thermal length scale $\xi(T)\propto T^{-1/4}$; within this length scale, orphan correlations are non-decaying. }
    \label{fig:SpinCorr}
\end{figure}

Fig. \ref{fig:SpinCorr} compares  effective theory with  Monte Carlo simulations for the correlations of two orphan spins as distance and temperature are varied for a  system  of $N=3200$ spins. We find very good agreement,  up to the thermal length $\xi(T)$, beyond which interactions are screened by thermal excitations.

From the perspective of the higher-rank gauge theory, the orphans correspond
to gauge charges (fractons), and  their constant interaction at long distances implies infinite range interactions between fractons.
Properties like this have already been predicted for fractonic spin liquids in
\cite{pretko_generalized_2017, pretko_subdimensional_2017}. 
At first glance, the infinite range interaction seems to suggest that the fractons are confined, since separating them to large distances requires an energy scaling with the distance. 
However,  energetics and dynamics in our setting work differently--the large free energy cost for creating far separated pairs has already been paid by the dilution itself, and fracton immobility prevents them collapsing back together.

\textit{Discussion \& outlook} 
We have analyzed the role of dilution by non-magnetic vacancies in a family of CSLs. 
We find that clusters of vacancies lead to the emergence
of effectively free (orphan) spins with a continuously tunable, and hence generically irrational, magnetic moment. Whether there is a relation of this phenomenon to irrational intrinsic charges of excitations in water and spin ice~\cite{MoeSon_2010} is an interesting open question. 
Irrational orphans can be expected more generally in CSLs involving varied contributions of spins to the constraints, or distinct energy scales enforcing the constraints, such as on the breathing pyrochlore lattice \cite{benton2015}.
This, along with the influence of quantum effects, will be explored in future work \cite{inprep}.

Focusing on a higher-rank spin liquid,
realizing a rank-2 $U(1)$ gauge theory, we find that the orphans induce non-decaying extended spin textures and {\it distance-independent} interactions in the low temperature limit. 
The underlying higher-rank gauge theory, endows the non-decaying spin textures with a characteristic quadrupolar angular dependence. 
These results illustrate the fundamentally distinct character of higher rank spin liquids, relative to more conventional CSLs, and how these can be revealed via the response to disorder.

\textit{Acknowledgements}--- This work was in part supported by the Deutsche
Forschungsgemeinschaft under Grants No. SFB 1143
(Project No. 247310070) and the cluster of excellence
ct.qmat (EXC 2147, Project No. 390858490).

\bibliography{Irrational_higher_gauge}

\clearpage

\renewcommand{\t}[1]{\text{#1}}
\renewcommand{\theequation}{S\arabic{equation}}
\renewcommand{\selectlanguage}[1]{}
\renewcommand{\thefigure}{S\arabic{figure}}

\setcounter{equation}{0}
\setcounter{secnumdepth}{1}

\renewcommand{\thesection}{S\arabic{section}}
\onecolumngrid
\begin{center}
  \textbf{\large Supplemental material for ``Irrational moments and signatures of higher-rank gauge theories in diluted classical spin liquids''}\\[.2cm]
  R. Flores-Calderon,$^{1,2,*}$ Owen Benton,$^{1,3}$ and Roderich Moessner$^{1}$\\[.1cm]
  {\itshape ${}^1$Max Planck Institute for Chemical Physics of Solids, Nöthnitzer Strasse 40, 01187 Dresden, Germany\\
  ${}^2$ Max Planck Institute for the Physics of Complex Systems, Nöthnitzer Strasse 38, 01187 Dresden, Germany\\
${}^3$School of Physical and Chemical Sciences, Queen Mary University of London, London, E1 4NS, United Kingdom}\\
  ${}^*$Electronic address: rflorescalderon@pks.mpg.de\\
(Dated: \today)\\[1cm]
\end{center}

\section{Orphan magnetization with the Lagrange multiplier method for the Honeycomb model}

In this section we will calculate the magnetic moment and effective Hamiltonian of a snowflake with just one spin left, refered to as an orphan spin. Instead of dealing with the full statistical field theory we will first consider a simple argument based on approximating the snowflake clusters as independent and working in the ground state manifold. Our starting point is the Hamiltonian of the undiluted honeycomb-snowflake model on the Honeycomb lattice. We consider $O(3)$ Heisenberg spins $\vec{S}$ in a magnetic field $\vec{h}$:
\begin{align}
    \mathcal{H}=\frac{J}{2} \sum_{\mathrm{hex}} \left(\sum_{i \in \mathrm{hex}} \vec{S}_i+\gamma \sum_{i \in\langle\mathrm{hex}\rangle} \vec{S}_i \right)^2-\sum_{i} \vec{S}_{i}\cdot\vec{h}=\frac{J}{2} \sum_{\mathrm{hex}}\vec{L}_i^2 -\sum_{i} \vec{S}_{i}\cdot\vec{h} \label{HoneycombHam}
\end{align}

\begin{figure}
    \centering
     \sidesubfloat[]{\includegraphics[width=0.3 \textwidth ]{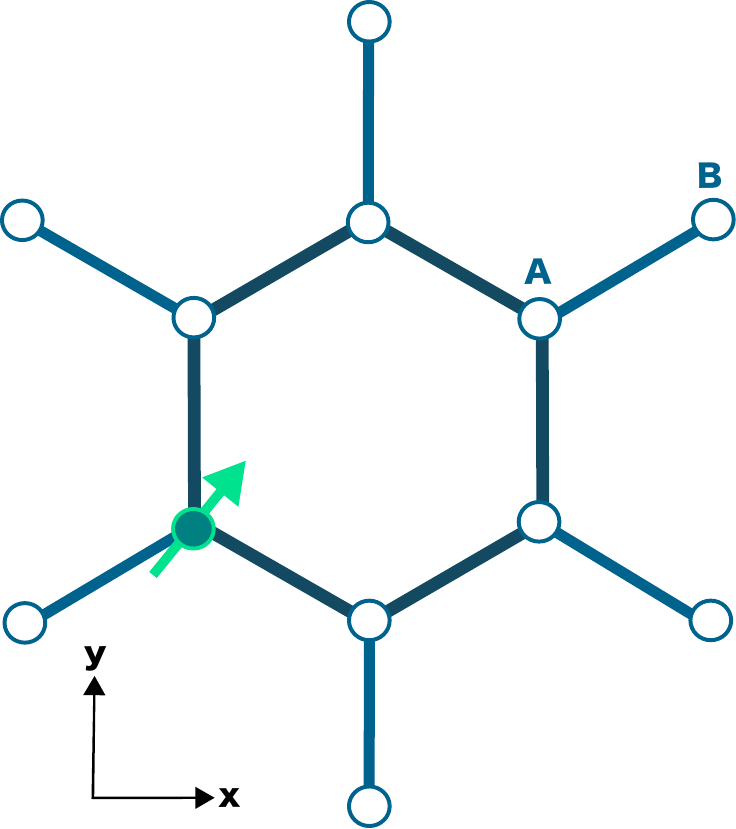}}
     \sidesubfloat[]{\includegraphics[width=0.3 \textwidth ]{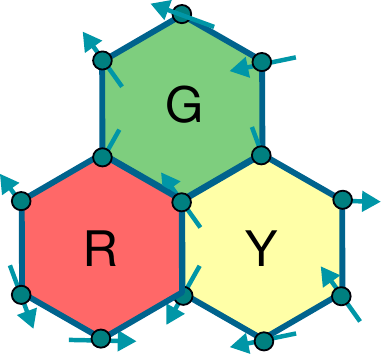}}
     \sidesubfloat[]{\includegraphics[width=0.3 \textwidth ]{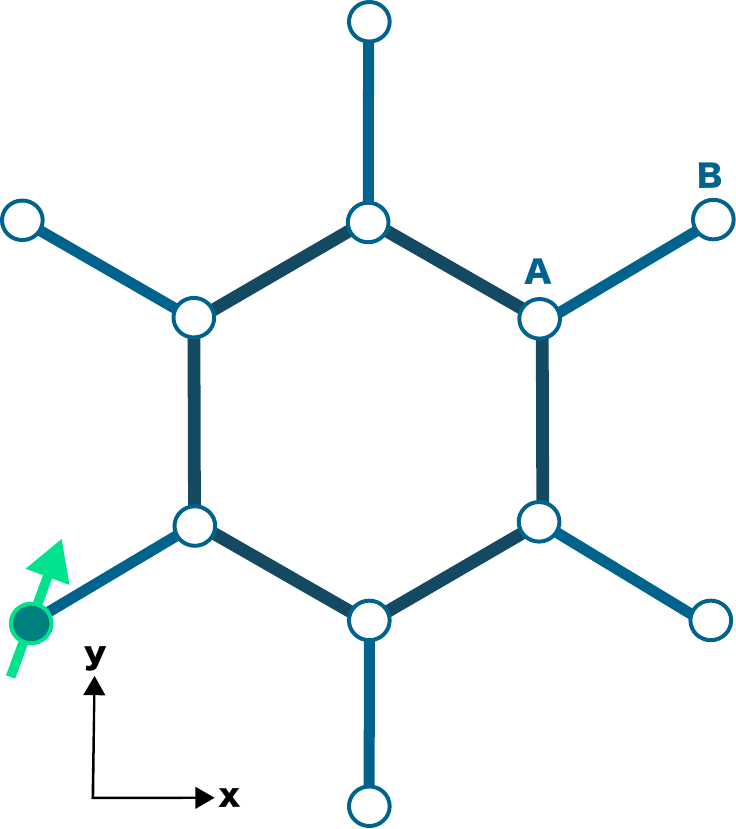}}
    \caption{Schematic of the honeycomb lattice orphan snowflake. a) The orphan spin depicted in green is of type $O_1$ since it stays inside the inner darker hexagon. In the schematic the spin is located on the $B$ sub-lattice. b) Coloring of the three types of hexagons used for the Lagrange multiplier construction.  c) An orphan spin of type $O_2$ is depicted localized on the $A$ sub-lattice.}
    \label{fig:OrphansHoney}
\end{figure}

The honeycomb lattice can be partitioned into three sets
of non-overlapping hexagons, which we label by $\lambda\in \{R,G,Y\}$ for red, green, yellow as illustrated in Fig. \ref{fig:OrphansHoney} (b). 
These hexagons are located at the three distinct sites of the dual triangular lattice of the initial Honeycomb.
Each spin participates in exactly two snowflakes coming
from each set of hexagons.
Thus summing over each subset of hexagons individually should yield the same result, since they each contain the same set of spins.
We now introduce Lagrange multipliers, $\vec{\mu}_1, \vec{\mu}_2$ to enforce this fact.
Let us call the vector of ground state constraints on each type of hexagon $\vec{L}_\lambda$ so that we can write:
\begin{align}
    \mathcal{H}=\frac{J}{2} \sum_{\lambda\in \{R,G,Y\}} \sum_{\mathrm{hex}(\lambda)} \vec{L}_\lambda^2-\sum_{i} \vec{S}_{i}\cdot\vec{h}+\vec{\mu}_1\left(\sum_{\mathrm{hex}(R)}\vec{L}_R-\sum_{\mathrm{hex}(G)}\vec{L}_G\right)+\vec{\mu}_2\left(\sum_{\mathrm{hex}(R)}\vec{L}_R-\sum_{\mathrm{hex}(Y)}\vec{L}_Y\right),
\end{align}
we can now use the fact that counting every spin separately or through the constraint vector should give the same result, when the repeated spins covered by the constraint vector are subtracted. Since each spin is covered three times by the first $i \in \mathrm{hex}$ sum and also three times for the boundary of the snowflake $i \in\langle\mathrm{hex}\rangle$ with an extra $\gamma$ factor we obtain:

\begin{align}
\sum_{\lambda\in \{R,G,Y\}} \sum_{\mathrm{hex}(\lambda)} \vec{L}_\lambda=3(1+\gamma)\sum_{i} \vec{S}_{i},
\end{align}
we use now this equation to get the Lagrange multipliers $\mu_{i}$ inside the first sum together with the magnetic field. They will enter with unknown coefficients $\beta^{\sigma}_\lambda$ so that we have the resulting Hamiltonian be (up to a constant):
\begin{align}
    \mathcal{H}=\frac{J}{2} \sum_{\lambda\in \{R,G,Y\}} \sum_{\mathrm{hex}(\lambda)} \left( \vec{L}_\lambda-\dfrac{\alpha}{J}\vec{h}-\dfrac{\beta^1_\lambda}{J}\vec{\mu}_1-\dfrac{\beta^2_\lambda}{J}\vec{\mu}_2\right)^2 .\label{HoneycombLagrange}
\end{align}
In order to match the previous Hamiltonian we must require then that 

\begin{align}
    \alpha =\dfrac{1}{3(1+\gamma)}\quad \beta^1_\lambda=\delta_{\lambda,R}-\delta_{\lambda,G},\quad  \beta^2_\lambda=\delta_{\lambda,R}-\delta_{\lambda,Y},\label{alphagamma}
\end{align}
it is clear now that the ground state is the one which satisfies:
\begin{align}
     \vec{L}_\lambda=\dfrac{\alpha}{J}\vec{h}+\dfrac{\beta^1_\lambda}{J}\vec{\mu}_1+\dfrac{\beta^2_\lambda}{J}\vec{\mu}_2,
\end{align}
which means the ground state magnetization is given by:
\begin{align}
     \vec{M}_{ud}=\sum_i \vec{S}_i=\alpha \sum_{\lambda\in \{R,G,Y\}} \sum_{\mathrm{hex}(\lambda)} \vec{L}_\lambda=\alpha \sum_{\lambda\in \{R,G,Y\}} \sum_{\mathrm{hex}(\lambda)} \dfrac{\alpha}{J}\vec{h}+\dfrac{\beta^1_\lambda}{J}\vec{\mu}_1+\dfrac{\beta^2_\lambda}{J}\vec{\mu}_2.
\end{align}
We must now determine the values of the Lagrange multipliers which are consistent with the constraints on the spins. One such constraint is that the sum over the $R$ hexagons of the constraint vector must be equal to the sum over the $G$ hexagons ,since they all describe the total number of spins. Mathematically this  means:
\begin{align}
    &\sum_{\mathrm{hex}(R)} \vec{L}_R= \dfrac{\alpha N }{3J}\vec{h}+\dfrac{N}{3 J}\vec{\mu}_1+\dfrac{N}{3J}\vec{\mu}_2=\dfrac{\alpha N }{3J}\vec{h}-\dfrac{N}{3 J}\vec{\mu}_1= \sum_{\mathrm{hex}(G)} \vec{L}_G\\
    &\sum_{\mathrm{hex}(R)} \vec{L}_R= \dfrac{\alpha N }{3J}\vec{h}+\dfrac{N}{3 J}\vec{\mu}_1+\dfrac{N}{3J}\vec{\mu}_2=\dfrac{\alpha N }{3J}\vec{h}-\dfrac{N}{3 J}\vec{\mu}_2= \sum_{\mathrm{hex}(Y)} \vec{L}_Y,
\end{align}
where $N$ is the total number of hexagons, we also repeated the constraint for the $Y$ honeycombs. To satisfy both equations simultaneously we see the only option is that $\mu_1=\mu_2=0$. So the undiluted ground state magnetization is given by:
\begin{align}
     \vec{M}_{ud}=\alpha \sum_{\lambda\in \{R,G,Y\}} \sum_{\mathrm{hex}(\lambda)} \dfrac{\alpha}{J}\vec{h}=\dfrac{N }{9J(1+\gamma)^2}\vec{h}
\end{align}

Let us proceed now to the diluted case of one orphan spin in a given snowflake which we take to be an $R$ hexagon. We have in this case different Lagrange multipliers which we distinguish by a prime. First let us take the spin to be in the inner hexagon and name this an $O_1$ orphan spin. The honeycomb lattice has two sub-lattices which we label $A,B$ and are pictured in Fig. \ref{fig:OrphansHoney}, we consider first the orphan spin to be in sub-lattice $B$ of the inner hexagon.

With this in mind the diluted ground state magnetization is given by:
\begin{align}
     &\vec{M}_{d}=\sum_i \vec{S}_i=\alpha \sum_{\lambda\in \{R,G,Y\}} \sum_{\mathrm{hex}(\lambda)} \vec{L}_\lambda=  \dfrac{\alpha N}{3} \left(\dfrac{\alpha }{ J}\vec{h}+\dfrac{1}{J}\vec{\mu}_1'+\dfrac{1}{J}\vec{\mu}_2'\right)+\dfrac{\alpha N}{3} \left(\dfrac{\alpha }{ J}\vec{h}-\dfrac{1}{J}\vec{\mu}'_1 \right)+\dfrac{\alpha N}{3} \left(\dfrac{\alpha }{ J}\vec{h}-\dfrac{1}{J}\vec{\mu}'_2 \right) \\
     &-\alpha \left(\dfrac{\alpha }{ J}\vec{h}+\dfrac{1}{J}\vec{\mu}'_1+\dfrac{1}{J}\vec{\mu}'_2 \right)+\alpha \vec{S}=\dfrac{(N-1) \alpha^2 }{ J}\vec{h}-\dfrac{\alpha }{ J}\left(\vec{\mu}'_1+\vec{\mu}'_2 \right)+\alpha \vec{S}\ ,
\end{align}
where we decomposed the magnetization in terms of the three types of hexagons and subtracted the constraint of the orphan snowflake. We need to substract it since the orphan snowflake has just one spin $\vec{S}$  contributing to the sum. Requiring again the same lattice constraint on the hexagons so that we obtain the equations:
\begin{align}
    &\sum_{\mathrm{hex}(R)} \vec{L}_R= \dfrac{ N}{3} \left(\dfrac{\alpha }{ J}\vec{h}+\dfrac{1}{J}\vec{\mu}_1'+\dfrac{1}{J}\vec{\mu}_2'\right) -\left(\dfrac{\alpha }{ J}\vec{h}+\dfrac{1}{J}\vec{\mu}'_1+\dfrac{1}{J}\vec{\mu}'_2 \right)+ \vec{S}=\dfrac{ N}{3} \left(\dfrac{\alpha }{ J}\vec{h}-\dfrac{1}{J}\vec{\mu}_1'\right) = \sum_{\mathrm{hex}(G)} \vec{L}_G\\
    &\sum_{\mathrm{hex}(R)} \vec{L}_R= \dfrac{ N}{3} \left(\dfrac{\alpha }{ J}\vec{h}+\dfrac{1}{J}\vec{\mu}_1'+\dfrac{1}{J}\vec{\mu}_2'\right) -\left(\dfrac{\alpha }{ J}\vec{h}+\dfrac{1}{J}\vec{\mu}'_1+\dfrac{1}{J}\vec{\mu}'_2 \right)+ \vec{S}=\dfrac{ N}{3} \left(\dfrac{\alpha }{ J}\vec{h}-\dfrac{1}{J}\vec{\mu}_2'\right) = \sum_{\mathrm{hex}(Y)} \vec{L}_Y
\end{align}
Simplifying this equation we obtain:
\begin{align}
    &  \dfrac{ 2 N-3}{3}\vec{\mu}_1'+\dfrac{ N-3}{3}\vec{\mu}_2'= \alpha \vec{h} - J\vec{S}\\
    &  \dfrac{ 2 N-3}{3}\vec{\mu}_2'+\dfrac{ N-3}{3}\vec{\mu}_1'= \alpha \vec{h} - J\vec{S}
\end{align}
Choosing the field to be $\vec{h}=h\hat{e}_z$ we obtain the solution:
\begin{align}
    \mu_1=\mu_2= \dfrac{h\alpha-JS}{N-2}\ ,
\end{align}
which implies the orphan magnetization at zero temperature is given by:
\begin{align}
    \vec{M}_{O_1}=\vec{M}_{d}-\vec{M}_{ud}=\left(\alpha S-\dfrac{\alpha^2}{J}h +\dfrac{\alpha^2 }{J (N-2)}(JS-h \alpha))\right)\hat{e}_z\ .
\end{align}
For the finite temperature result we can go back to the Hamiltonian of eq. \eqref{HoneycombHam} which becomes eq. \eqref{HoneycombLagrange} with the magnetic field included. We then see that in the thermodynamic limit and thinking each snowflake as independent, the only term coupling the orphan spin is the one with the magnetic field so that :
\begin{align}
    H_{O_1}=-\alpha \ \vec{S}\cdot \vec{h}
\end{align}
From this approximate Hamiltonian we can obtain the average magnetization of the orphan spin as that of a free spin, with a modified magnetic moment, in terms of the Langevin function $M_{\alpha S}(h,T)= \alpha S \ L \left( \alpha S \dfrac{h}{T} \right),\ L(x)=\coth(x)-1/x$ for a Hamiltonian of the usual form $-\vec{h}\cdot \vec{S}$.

\section{Hybrid field theory for the honeycomb-snowflake model}

Although the previous calculation captures the basic response of the orphan spin, a better result can be obtained by using the large $N$ limit. This relies on assuming that the number of components for each spin $N$ is large enough that the self-consistent Gaussian approximation is valid. Physically we assume the fixed spin length constraint to be satisfied only on average. We do this by introducing a Lagrange multiplier $\rho_{\vec{r}}$ to fix the expectation value of the magnitude of the new \textit{soft spins}, following  Ref. \cite{sen_vacancy-induced_2012}. This gives the effective partition function for the undiluted spin liquid:

\begin{align}
Z_{\mathrm{eff}} ^{\text{ud}}& =\int \mathcal{D} \vec{\phi} \exp \left(-S_{\mathrm{eff}}^{\text{ud}}\right) \\
S_{\mathrm{eff}}^{\text{ud}} & =\frac{1}{2} \sum_{\vec{r}} \rho_{\vec{r}} \vec{\phi}_{\vec{r}}^2+\frac{1}{T} \mathcal{H}\left(\left\{\vec{\phi}_{\vec{r}}\right\}\right)=\frac{1}{2} \sum_{\vec{r}} \rho_{\vec{r}} \vec{\phi}_{\vec{r}}^2+\frac{\beta J}{2} \sum_{\mathrm{hex}} \left(\sum_{i \in \mathrm{hex}} \vec{\phi}_i+\gamma \sum_{i \in\langle\mathrm{hex}\rangle} \vec{\phi}_i \right)^2-\beta\sum_{\vec{r}} \vec{\phi}_{\vec{r}}\cdot\vec{h}
\end{align}
,where the $\rho_{\vec{r}}$ stiffness coefficients are fixed by requiring $\expval{\vec{\phi}_{\vec{r}}^2}=S^2$. Since the spins are symmetry equivalent we can fix $\rho_{\vec{r}}=\rho$ which for $T\rightarrow 0,h\rightarrow 0,\gamma\rightarrow 0$ fixes it to be $\rho\approx 1.45 S^2$. We now Fourier Transform the quadratic part of the effective action by defining $\vec{\phi}_{\vec{r}}=\vec{\Phi}^\sigma_{\vec{R}_i}=\frac{1}{\sqrt{N}}\sum_{\vec{q}}\vec{\Phi}^{\sigma}_{\vec{q}} e^{-i\vec{q} \cdot \vec{r}}$, where $\sigma=A,B$ labels the sub-lattice in the Honeycomb unit cell and $\vec{R}_i$ is the vector pointing to the center of the hexagon. Each site of the snowflake has a given position $\vec{r}=\vec{R}_i+\vec{\delta}_l$ where $\delta_l$ labels the position of sites in the honeycomb or on the boundary of the snowflake. Fourier transforming to momentum space we obtain:

\begin{align}
S_{\mathrm{eff}}^{\text{ud}} & =\frac{1}{2} \rho \sum_{\vec{q},\sigma} \vec{\Phi}^{\sigma}_{\vec{q}}\cdot \vec{\Phi}^{\sigma}_{-\vec{q}} +\dfrac{\beta J}{2} \sum_{\vec{q},\sigma,\sigma'}\tilde{h}_{\sigma \sigma'}(\vec{q})\vec{\Phi}^{\sigma}_{-\vec{q}}\cdot \vec{\Phi}^{\sigma'}_{\vec{q}}-\beta L \sum_\sigma\vec{\Phi}^{\sigma}_{\vec{q}=0}\cdot \vec{h},
\end{align}
so that now if we measure $\beta$ in units of $JS^2$ ,the magnetic field in units of $J S$ together with $\phi \rightarrow S \phi$ and we have $L^2$ unit cells, then the stiffness coefficients now satisfy $\expval{\vec{\phi}_{\vec{r}}^2}=1$. The action of the undiluted Gaussian system with a magnetic field reads:
\begin{align}
S_{\mathrm{eff}}^{\text{ud}} & =\frac{1}{2}  \sum_{\vec{q},\sigma,\sigma'}\vec{\Phi}^{\sigma}_{-\vec{q}}\cdot h_{\sigma \sigma'}(\vec{q}) \vec{\Phi}^{\sigma'}_{\vec{q}}-\beta L \sum_\sigma\vec{\Phi}^{\sigma}_{\vec{q}=0}\cdot \vec{h},
\end{align}
 where the quadratic part of the action now contains also the stiffness condition so that in matrix form it is explicitly given by:
{\small
\begin{align}
    h(\vec{q}) = &\beta \begin{pmatrix}
     d_A+\frac{\rho}{\beta} && d_{AB} \\ d_{AB}^* && d_A+\frac{\rho}{\beta}
    \end{pmatrix}\quad d_{AB}= e^{-\frac{4}{3} i (q_x+q_y)}\left(e^{i q_x}+e^{i q_y}+e^{i (q_x+q_y)}+\gamma+\gamma e^{2 i q_x} +\gamma e^{2 i q_y} \right)^2 \notag \\ &d_{A}=3+3 \gamma^2 +2(1+2\gamma) (\cos(q_x) +\cos(q_y)+\cos(q_x-q_y))+2 \gamma^2 (\cos(2 q_x) +\cos(2 q_y)+\cos(2 q_x-2 q_y) ) \notag  \\
    &\hspace{1cm}  +2\gamma (\cos(q_x+q_y)+\cos(q_x-2 q_y)+\cos(2 q_x-q_y)).
\end{align}}%
The orphan spins can be treated now by replacing all field components at the vacancy sites for zero, while fixing exactly the length of the orphan spin. In this way we have a hybrid field theory, where the orphan spin interacts with the spin liquid background. This means the diluted system has a partition function of the form:
\begin{align}
    Z_{\mathrm{eff}}& =\int \mathcal{D} \vec{\phi} \int \mathcal{D} \vec{n}  \exp \left(-S_{\mathrm{eff}}^{\text{ud}}\right)\prod_\alpha\delta(\phi_{\vec{r}_o}^\alpha-n^\alpha)\prod_{\vec{r}_v,\alpha}\delta(\phi_{\vec{r}_v}^\alpha),
\end{align}
where $\alpha$ runs over the spin components, $\vec{r}_o$ indicates the orphan spin position and $\vec{r}_v$ runs over all the vacancy sites.  The delta functions make the dilution constraint exact, while the surrounding liquid is treated only by an averaged constraint.  We use now the representation of the Dirac delta distribution by an exponential integral to obtain:
\begin{align}
    Z_{\mathrm{eff}}& =\int \mathcal{D} \vec{n} \int \mathcal{D} \vec{\mu} \int \mathcal{D} \vec{\lambda} \int \mathcal{D} \vec{\Phi}  \exp \left(-S_{\mathrm{eff}}^{\text{ud}}\right) \exp{i\sum_{\vec{r}_v}\vec{\lambda}_{\vec{r}_v}\cdot\vec{\Phi}_{\vec{r}_v}^{\sigma(v)}+i \sum_{\vec{r}_o} \vec{\mu}_{\vec{r}_o}\cdot( \vec{\Phi}_{\vec{r}_o}^{\sigma(o)}-\vec{n}_{\vec{r}_o})}
\end{align}
where the integral measures are $ \mathcal{D} \vec{n}=\prod_{\vec{r}_o,\alpha} \dd n^\alpha_{\vec{r}_o} \delta(\vec{n}^2_{\vec{r}_o}-1),\mathcal{D} \vec{\lambda}= \prod_{\vec{r}_v,\alpha} \dd \lambda_{\vec{r}_v,\alpha}^\alpha/2\pi,\mathcal{D} \vec{\lambda}= \prod_{\vec{r}_o,\alpha} \dd \mu_{\vec{r}_o}^\alpha /2\pi$ and the notation $\sigma(v)$ denotes the sub-lattice where the $\vec{r}_v$ vacancy is located. Thanks to the linear coupling of the $\vec{\phi}$ field to the other Lagrange multiplier fields and magnetic field one can perform the Gaussian integrals of $\vec{\phi}$ exactly, the general integration of Gaussian fields requires to define the total current coming from the linear part of the action, which is:

\begin{align}
    S_J= -\sum_{\vec{q},\sigma}\vec{J}_{\vec{q}\sigma}\cdot \vec{\Phi}^\sigma_{\vec{q}}=-i \sum_{\vec{q},\sigma}\sum_{\vec{r}_v\in \sigma}\dfrac{1}{L}\vec{\lambda}_{\vec{r}_v}e^{-i\vec{q}\cdot \vec{r}_v} \vec{\Phi}_{\vec{q}}^\sigma -i \sum_{\vec{q},\sigma}\sum_{\vec{r}_o\in \sigma}\dfrac{1}{L}\vec{\mu}_{\vec{r}_o}e^{-i\vec{q}\cdot \vec{r}_o} \vec{\Phi}_{\vec{q}}^\sigma-\beta L \sum_{\vec{q},\sigma}\delta_{\vec{q},0}\vec{h}\cdot\vec{\Phi}^{\sigma}_{\vec{q}}.
\end{align}
So that one can write the diluted partition function and effective action as:
\begin{align}
    Z_{\mathrm{eff}}& =\int \mathcal{D} \vec{n} \int \mathcal{D} \vec{\mu} \int \mathcal{D} \vec{\lambda} \exp{-i \sum_{\vec{r}_o} \vec{\mu}_{\vec{r}_o}\cdot\vec{n}_{\vec{r}_o}} \int \mathcal{D} \vec{\Phi}  \exp \left(-S_{\mathrm{eff}}^0\right) \exp{\sum_{\vec{q},\sigma}\vec{J}_{\vec{q}\sigma}\cdot \vec{\Phi}^\sigma_{\vec{q}}},\\
    S_{\mathrm{eff}}^0&=\frac{1}{2}  \sum_{\vec{q},\sigma,\sigma'}\vec{\Phi}^{\sigma}_{-\vec{q}}\cdot h_{\sigma \sigma'}(\vec{q}) \vec{\Phi}^{\sigma'}_{\vec{q}} , \quad \vec{J}_{\vec{q}\sigma}=i \sum_{\vec{r}_v\in \sigma}\dfrac{1}{L}\vec{\lambda}_{\vec{r}_v}e^{-i\vec{q}\cdot \vec{r}_v}  +i \sum_{\vec{r}_o\in \sigma}\dfrac{1}{L}\vec{\mu}_{\vec{r}_o}e^{-i\vec{q}\cdot \vec{r}_o} +\beta L \delta_{\vec{q},0}\vec{h}.
\end{align}
From this equation we have at this level four fields to integrate over,  two come from Lagrange multipliers enforcing constraints, one from fixing the orphan spin length and one describing the surrounding spin liquid. The  former fields act as sources for the spin liquid and enter through the vector current.  We proceed now by integrating out the spin liquid background. The integral over the $\Phi$ field is a Gaussian integral, but care must be taken in restricting to a real $\phi(x)$ , which means the measure must take into account the real field conditions $\Phi_q=-\Phi_{-q}^*$ and $h^T(\vec{q})=h(-\vec{q})$, we obtain then, effectively:
\begin{align}
     &Z_{\mathrm{eff}} =\int \mathcal{D} \vec{n} \int \mathcal{D} \vec{\mu} \int \mathcal{D} \vec{\lambda} \exp{-i \sum_{\vec{r}_o} \vec{\mu}_{\vec{r}_o}\cdot\vec{n}_{\vec{r}_o}}\exp{\dfrac{1}{2}\sum_{\vec{q},\sigma,\sigma'}\vec{J}_{\vec{q}\sigma}h^{-1}_{\sigma\sigma'}(\vec{q})\vec{J}_{-\vec{q}\sigma}}=\int \mathcal{D} \vec{n} \int \mathcal{D} \vec{\mu} \int \mathcal{D} \vec{\lambda} e^{-S_{\mathrm{eff}}^{\mu\lambda}\left[ \vec{n}\right]},\\
     &S^{\mu\lambda}_{\mathrm{eff}}=-\dfrac{1}{2}\sum_{\vec{q},\sigma,\sigma'}\vec{J}_{\vec{q}\sigma}h^{-1}_{\sigma\sigma'}(\vec{q})\vec{J}_{-\vec{q}\sigma'}+i \sum_{\vec{r}_o} \vec{\mu}_{\vec{r}_o}\cdot\vec{n}_{\vec{r}_o}.
\end{align}
To proceed with the calculation, let us define now a generating functional which will be useful for obtaining diluted correlation functions of the orphan spin:
\begin{align}
    Z_{\mathrm{eff}} [\tilde{J},h]=\int \mathcal{D} \vec{n} \int \mathcal{D} \vec{\mu} \int \mathcal{D} \vec{\lambda}\exp{\dfrac{1}{2}\sum_{\vec{q},\sigma,\sigma'}(J^\alpha_{\vec{q}\sigma}+\tilde{J}^{\alpha}_{\vec{q}\sigma})h^{-1}_{\sigma\sigma'}(\vec{q})(J^\alpha_{-\vec{q}\sigma'}+\tilde{J}^{\alpha}_{-\vec{q}\sigma'})-i \sum_{\vec{r}_o} \vec{\mu}_{\vec{r}_o}\cdot\vec{n}_{\vec{r}_o}}.
\end{align}
Here the vector field $\tilde{J}$  acts as the external source, which we can make use of for calculating moments of the distribution, thus it must enter in the action in the same way  the physical current vector $J$ does . To proceed let us rename the vacancy sites and combine them with the orphan spin site by defining the sites in the B sublattice to be $\vec{x}_i^B$ with $\vec{x}_1^B$ the orphan spin site and $i=1,\dots,6$ gives the vacancy site positions in the unit cell. Similarly $\vec{x}_i^A$ gives the sites of the vacancies present in the A sublattice. We can then combine the Lagrange multipliers from the orphan spin and vacancies into a single multidimensional object $\Lambda_i^{\alpha\sigma}$ defined by:
\begin{align}
    \Lambda_i^{\alpha\sigma}=\lambda^\alpha_{\vec{r}_v}\delta_{\vec{x}^\sigma_i,\vec{r}_v}+\mu^\alpha_{\vec{r}_o}\delta_{\vec{x}^\sigma_i,\vec{r}_o}.
\end{align}
We will need the undiluted correlation matrix :
\begin{align}
    C_{\sigma,\sigma'}^{i-j}=\dfrac{1}{N} \sum_{\vec{q}}h^{-1}_{\sigma\sigma'}(\vec{q}) e^{-i(\vec{x}_i^\sigma-\vec{x}_j^{\sigma'})\cdot\vec{q}}=\dfrac{1}{3}\expval{\vec{\phi}(\vec{x}_i^{\sigma})\cdot \vec{\phi}(\vec{x}_j^{\sigma'})}_{ud}.
\end{align}
We make use of the momentum space representation for convenience. It is worth noting that when summed over all the sublattices this representation gives rise to the spin-spin structure factor $\expval{\vec{S}(\vec{q})\cdot \vec{S}(-\vec{q}))}$, which characterizes the spin liquid state. The effective action becomes before integrating the $\Lambda$ fields:
\begin{align}
    &S_{\mathrm{eff}}^{\Lambda}= -\dfrac{1}{2N}\sum_{\vec{q},\sigma,\sigma'}(i\sum_{n}\Lambda_n^{\sigma\alpha}e^{-i \vec{q}\cdot \vec{x}_n^{\sigma}}+\beta N \delta_{q,0}h^\alpha+ \tilde{J}^{\alpha}_{\vec{q}\sigma})h^{-1}_{\sigma\sigma'}(\vec{q})(i\sum_{m}\Lambda_m^{\sigma'\alpha}e^{i \vec{q}\cdot \vec{x}_m^{\sigma'}}+\beta N \delta_{-q,0}h^\alpha+ \tilde{J}^{\alpha}_{-\vec{q}\sigma'})+i \sum_{\alpha} \Lambda^{B\alpha}_1 n^\alpha\\
    &= \dfrac{1}{2}\sum_{n,m,\sigma\sigma'}\Lambda_n^{\alpha\sigma}\Lambda_m^{\alpha\sigma'} C_{\sigma,\sigma'}^{n-m}- \sum_{n,\sigma\sigma'}\Lambda_n^{\alpha\sigma}\left(i \beta h^\alpha h^{-1}_{\sigma\sigma'}(0)+i \sum_{\vec{q}} \dfrac{1}{N}\tilde{J}^{\alpha}_{-\vec{q}\sigma'}e^{-i \vec{q}\cdot \vec{x}_n^{\sigma}}h^{-1}_{\sigma\sigma'}(\vec{q})-i\delta_{n,1}\delta_{\sigma,B}n^\alpha\right)+S_1\left[h,\tilde{J}\right]\\
    &S_1\left[h,\tilde{J}\right] = -\dfrac{1}{2}\beta^2 N^2 h^2 \sum_{\sigma \sigma'}h^{-1}_{\sigma\sigma'}(0) - \beta h^\alpha \sum_{\sigma \sigma'}\tilde{J}^{\alpha}_{0\sigma}h^{-1}_{\sigma\sigma'}(0)-\dfrac{1}{2 N} \sum_{\sigma \sigma'}\tilde{J}^{\alpha}_{\vec{q}\sigma}h^{-1}_{\sigma\sigma'}(\vec{q}))\tilde{J}^{\alpha}_{-\vec{q}\sigma'}
\end{align}
This action is again quadratic in the $\Lambda$ fields, the currents come now from the spin liquid which has been integrated out so that it sources now the Lagrange multiplier fields together with the fixed orphan spin length condition.  The last contribution to the action, $S_1$ does not depend on the orphan spin and comes only from the external sources $h,\tilde{J}$ . It is worth noting that for $\tilde{J}=0$ we have only a quadratic contribution of the magnetic field in $S_1$. As such it will not affect the magnetization, which involves first order derivatives of the magnetic field and we ignore it in the following. We proceed now to integrate them out to obtain:   
\begin{align}
    Z_{\mathrm{eff}} [0,h]&= \int \mathcal{D} \vec{n}\exp{\dfrac{1}{2} (\sum_{\sigma'}i\beta h^\alpha h^{-1}_{\sigma\sigma '}(0)-i \delta_{n,1}\delta_{\sigma,B}n^\alpha) ((\mathcal{P}_OC\mathcal{P}_O)^{-1})_{\sigma \tilde \sigma}^{n,m} (\sum_{\sigma'}i\beta h^\alpha h^{-1}_{\tilde{\sigma}\sigma '}(0)-i \delta_{m,1}\delta_{\tilde{\sigma},B}n^\alpha) 
-\tilde{S}_0[h] }\\
&=e^{-S_0[h]}\int \mathcal{D} \vec{n}\exp{\beta n^\alpha h^\alpha \sum_{\sigma',\sigma}  h^{-1}_{\sigma\sigma '}(0) (C^{-1})_{\sigma B}^{n,1}}=Z_{0}Z_{O_1},
\end{align}
where we used the fact that $ ((\mathcal{P}_OC\mathcal{P}_O)^{-1})_{\sigma \tilde \sigma}^{n,m}= ((\mathcal{P}_OC\mathcal{P}_O)^{-1})_{\tilde \sigma \sigma }^{m,n}$, the notation means we project the correlation matrix to the orphan spin sites and then invert the matrix afterwards. We also collected all terms independent of $n$ inside the action $S_0$ and took  $\tilde{J}=0$  since we focus on the magnetization. To find the diluted magnetization we now take the derivative with respect to $h_z$, specializing to a perpendicular magnetic field. From the form of the action we see two contributions:
\begin{align}
    M^\mu&=\dfrac{1}{\beta Z_{\mathrm{eff}} [0,h]}\dfrac{\delta Z_{\mathrm{eff}} [0,h]}{\delta h^\mu}=\dfrac{1}{\beta Z_{0}Z_{O_1}}e^{-S_0[h]}\int \mathcal{D} \vec{n} \ \beta (\alpha n^\mu)e^{\beta (\alpha \vec{n})\cdot\vec{h}}+\dfrac{Z_{O_1}}{\beta Z_{0}Z_{O_1}}\dfrac{\delta Z_{0}}{\delta h^\mu}.
\end{align}
The first contribution is proportional to the orphan spin vector, while the second one has the response of the surrounding undiluted spin liquid. Clearly the dilution effect is encoded in the first term only, let us define then the orphan magnetization with corresponding magnetic moment $\alpha$ as:
\begin{align}
   \vec{M}_{O_1} &\equiv \dfrac{1}{Z_{O_1}}\int \mathcal{D} \vec{n} \ (\alpha \vec{n})\ e^{\beta (\alpha \vec{n})\cdot\vec{h}},\qquad \alpha\equiv \sum_{\sigma',\sigma,n}  h^{-1}_{\sigma\sigma '}(0) ((\mathcal{P}_OC\mathcal{P}_O)^{-1})_{\sigma B}^{n,1}.
\end{align}
In this representation the orphan spin behaves like a free spin within a magnetic field $\vec{h}$ , but with an emergent magnetic moment $\alpha$ whose origin comes directly from the correlations of the surrounding spin liquid. This agrees indeed with the naive argument of the last section, furthermore the exact parameter dependence of eq. \eqref{alphagamma} matches, once further approximations are done as shown next.
 
\section{Texture induced by orphan spin and long range approximation of the orphan spin magnetic moment}

\subsection{Derivation from the hybrid-field theory in the long-range limit}

We next analyze the resulting spin texture, i.e., the configuration around an orphan spin, by considering the spatial distribution of the expectation value of the $z$ component of spin  $\expval{S^z_{\vec{r}}}$. In the field theory, this can be calculated by imposing the orphan spin constraint as well as fixing the spin at the measuring location so as to integrate out all other degrees of freedom. As an approximation to the spin texture at the unit cell $\vec{R}_2$ away from the orphan spin located in the unit cell $\vec{R}_1$, we may choose to impose only the sum of the orphan snowflake to be equal to the orphan spin. Thus, instead of exactly writing out the vacancies and orphan spin location, we assume the detailed internal structure should not matter  far away from the orphan snowflake:
\begin{align}
    Z_{\text{text}}[\tilde{J}_z]&= \int \mathcal{D} \vec{\phi}\int \mathcal{D} \vec{n}_1 \ \delta(\vec{\phi}^1_\gamma-\vec{n}_1 )\int \mathcal{D} \vec{n}_2  \ \delta(\vec{\phi}_{\vec{r}_2}-\vec{n}_2) \ e^{-S_{\text{ud}}[\vec{\phi},\vec{h}]+\tilde{J}_z n_2^z} \\
    &= \int \mathcal{D} \vec{\phi}\int \mathcal{D} \vec{n}_1 \int \mathcal{D} \vec{n}_2 \int \mathcal{D} \vec{\mu}_1 \int \mathcal{D} \vec{\mu}_2 \exp{i(\vec{\phi}^1_\gamma-\vec{n}_1 )\cdot \vec{\mu}_1+i(\vec{\phi}_{\vec{r}_2}-\vec{n}_2)\cdot\vec{\mu}_2+\tilde{J}_z n_2^z-S_{\text{ud}}},
\end{align}
where we have defined the snowflake vector located in the unit cell $\vec{R}_1$ as $\vec{\phi}^1_\gamma=\sum_{i \in \mathrm{hex}} \vec{\phi}_i^1+\gamma \sum_{i \in\langle\mathrm{hex}\rangle} \vec{\phi}_i^1 $. We have also introduced a generating field $\tilde{J}_z$ so as to be able to take derivatives and calculate texture later.  Since the diluted snowflake has one spin, the orphan spin,  we must have  $\vec{\phi}^1_\gamma=\vec{n}_1$ , which is taken into account in the first Dirac Delta. The second Dirac Delta fixes the measuring site spin length, but it will not be needed to arrive at a first order expression. In the last step we have used again the Dirac Delta representation in terms of an exponential integral to get an effective action. To make contact with the last section we rewrite the action in terms of the Fourier basis:
\begin{align}
    i\vec{\phi}^1_\gamma \cdot \vec{\mu}_1 &= \dfrac{i}{\sqrt{N}}\left(\sum_{\vec{q}}\sum_{i \in \mathrm{hex}} \vec{\Phi}_{\vec{q}}^{\sigma(i)}e^{-i\vec{q}\cdot \vec{r}(i)}+\gamma \sum_{i \in\langle\mathrm{hex}\rangle} \vec{\Phi}_{\vec{q}}^{\sigma(i)}e^{-i\vec{q}\cdot \vec{r}(i)}\right) \cdot \vec{\mu}_1
    =\dfrac{i}{\sqrt{N}}\left( \sum_{\vec{q},\sigma} \sum_{\vec{r}_h \in \sigma } \vec{\Phi}_{\vec{q}}^{\sigma}e^{-i\vec{q}\cdot \vec{r}_h }+ \gamma \sum_{\vec{r}_{\expval{h}} \in \sigma } \vec{\Phi}_{\vec{q}}^{\sigma}e^{-i\vec{q}\cdot \vec{r}_{\expval{h}} } \right) \cdot \vec{\mu}_1\\
    &=  \sum_{\vec{q},\sigma} \vec{J}^{\gamma}_{\vec{q}\sigma} \vec{\Phi}_{\vec{q}}^{\sigma}, \qquad  \vec{J}^{\gamma}_{\vec{q}\sigma} = \dfrac{i}{\sqrt{N}} \left(\sum_{\vec{r}_h \in \sigma } e^{-i\vec{q}\cdot \vec{r}_h }+ \gamma \sum_{\vec{r}_{\expval{h}} \in \sigma } e^{-i\vec{q}\cdot \vec{r}_{\expval{h}} } \right)\vec{\mu}_1,
\end{align}
where we label the sites in the inner hexagon of the orphan snowflake by $\vec{r}_h $ and the snowflake boundary by $\vec{r}_{\expval{h}}$. This equation again shows us how the Lagrange multiplier fields source, by giving rise to currents, the underlying spin liquid. The remaining contribution to the diluted action comes from the location where we want to measure the Zeeman spin texture:
\begin{align}
    i\vec{\phi}_{\vec{r}_2} \cdot \vec{\mu}_2 = \dfrac{i}{\sqrt{N}}  \sum_{\vec{q}} \vec{\Phi}_{\vec{q}}^{\sigma_2}  e^{-i\vec{q}\cdot \vec{r}_2 } \cdot \vec{\mu}_2=\dfrac{i}{\sqrt{N}}  \sum_{\vec{q},\sigma} \delta_{\sigma,\sigma_2}\vec{\Phi}_{\vec{q}}^{\sigma}  e^{-i\vec{q}\cdot (\vec{R}_2+\vec{\tau}_{\sigma})} \cdot \vec{\mu}_2 \ .
\end{align}
We will now group together the two contributions to the current coming from the two Lagrange multiplier fields which give rise to an effective action and total vector current defined by:
\begin{align}
    &S_{\text{text}}=-i(\vec{\phi}^1_\gamma-\vec{n}_1 )\cdot \vec{\mu}_1-i(\vec{\phi}_{\vec{r}_2}-\vec{n}_2)\cdot\vec{\mu}_2+S_{\text{ud}}-\tilde{J}_z n_2^z=\frac{1}{2}  \sum_{\vec{q},\sigma,\sigma'}\vec{\Phi}^{\sigma}_{-\vec{q}}\cdot h_{\sigma \sigma'}(\vec{q}) \vec{\Phi}^{\sigma'}_{\vec{q}}-\sum_{\vec{q},\sigma} \vec{J}_{\vec{q}\sigma} \vec{\Phi}_{\vec{q}}^{\sigma}+i(\vec{n}_1\cdot \vec{\mu}_1+\vec{n}_2\cdot \vec{\mu}_2)-\tilde{J}_z n_2^z,\\
    &\vec{J}_{\vec{q}\sigma}=\beta \sqrt{N} \delta_{\vec{q},0} \vec{h}+i\sum_{\vec{r}_h \in \sigma } \dfrac{1}{\sqrt{N}}e^{-i\vec{q}\cdot \vec{r}_h }\vec{\mu}_1+ i\dfrac{1}{\sqrt{N}}\gamma \sum_{\vec{r}_{\expval{h}} \in \sigma } e^{-i\vec{q}\cdot \vec{r}_{\expval{h}} } \vec{\mu}_1+\dfrac{i}{\sqrt{N}}  \delta_{\sigma,\sigma_2} e^{-i\vec{q}\cdot (\vec{R}_2+\vec{\tau}_{\sigma})} \vec{\mu}_2\ .
\end{align}
This effective action is again quadratic in the $\vec{\phi}$ fields, which implies we can integrate them out so as to obtain now a partition function depending only on the Lagrange multiplier fields and the unit vectors:
\begin{align}
    &Z_{\text{text}}[\tilde{J}_z]= \int \mathcal{D} \vec{n}_1 \int \mathcal{D} \vec{n}_2 \int \mathcal{D} \vec{\mu}_1 \int \mathcal{D} \vec{\mu}_2 \exp{\dfrac{1}{2}\sum_{\vec{q},\sigma,\sigma'}\vec{J}_{\vec{q}\sigma}h^{-1}_{\sigma\sigma'}(\vec{q})\vec{J}_{-\vec{q}\sigma'}-i(\vec{n}_1\cdot \vec{\mu}_1+\vec{n}_2\cdot \vec{\mu}_2)+\tilde{J}_z n_2^z}\,\\
    &S_{\text{text}}^{\text{eff}}=-\dfrac{1}{2 N}\sum_{\vec{q},\sigma,\sigma'}\left(\beta N\delta_{\vec{q},0} \vec{h}+i A_{\sigma}(\vec{q})\vec{\mu}_1+i \delta_{\sigma,\sigma_2} e^{-i\vec{q}\cdot (\vec{R}_2+\vec{\tau}_{\sigma})} \vec{\mu}_2\right)h^{-1}_{\sigma\sigma'}(\vec{q})\left(\beta N\delta_{-\vec{q},0} \vec{h}+i A_{\sigma'}(-\vec{q})\vec{\mu}_1+i \delta_{\sigma',\sigma_2} e^{i\vec{q}\cdot (\vec{R}_2+\vec{\tau}_{\sigma'})} \vec{\mu}_2\right)\notag\\
    &+i(\vec{n}_1\cdot \vec{\mu}_1+\vec{n}_2\cdot \vec{\mu}_2)-\tilde{J}_z n_2^z.
\end{align}
Here we expanded the total current and defined a new function, $A_{\sigma}(\vec{q})$ useful for further computations: 
\begin{align}
    A_{\sigma}(\vec{q})=\sum_{\vec{r}_h \in \sigma } e^{-i\vec{q}\cdot \vec{r}_h }+ \gamma \sum_{\vec{r}_{\expval{h}} \in \sigma } e^{-i\vec{q}\cdot \vec{r}_{\expval{h}} }.
\end{align}
This is just the Fourier transform of the constraint vector and is important for characterizing topological defects in momentum space as noted before in ref. \cite{benton_topological_2021}, this defects give rise to pinch points in the structure factor and thus characterize the spin liquid state. Let us further rewrite the effective action in a matrix  structure so as to integrate the $\mu^\alpha_i$ field:
\begin{align}
    S_{\text{text}}^{\mu}&=\dfrac{1}{2 N}\begin{pmatrix}
        \mu_1^\nu  && \mu_2^\nu
    \end{pmatrix}
    \begin{pmatrix}
        \sum_{\vec{q},\sigma,\sigma'} A_{\sigma}(\vec{q})h^{-1}_{\sigma\sigma'}(\vec{q})A_{\sigma'} (-\vec{q}) && \sum_{\vec{q},\sigma} A_{\sigma}(\vec{q}) e^{i\vec{q}\cdot (\vec{R}_2+\vec{\tau}_{\sigma_2})}h^{-1}_{\sigma\sigma_2}(\vec{q})\\
        \sum_{\vec{q},\sigma} A_{\sigma}(\vec{q}) e^{i\vec{q}\cdot (\vec{R}_2+\vec{\tau}_{\sigma_2})}h^{-1}_{\sigma\sigma_2}(\vec{q}) &&  \sum_{\vec{q}}h^{-1}_{\sigma_2\sigma_2}(\vec{q})
    \end{pmatrix}\begin{pmatrix}
        \mu_1^\nu  \\ \mu_2^\nu
    \end{pmatrix}-\vec{J}_1\cdot \vec{\mu}_1-\vec{J}_2\cdot \vec{\mu}_2,\\
    \vec{J}_1&=i\beta \vec{h}\sum_{\sigma,\sigma'}h^{-1}_{\sigma\sigma'}(0)A_{\sigma'}(0) -i\vec{n}_1=\frac{2i\beta \vec{h}\ 3(1+\gamma)}{18 \beta  (\gamma +1)^2+\rho }-i\vec{n}_1,\\
    \vec{J}_2&=i\beta \vec{h}\sum_{\sigma}h^{-1}_{\sigma\sigma_2}(0)-i\vec{n}_2=\frac{i\beta \vec{h}}{18 \beta  (\gamma +1)^2+\rho }-i\vec{n}_2.
\end{align}
The action thus decomposes into sectors relating only the orphan snowflake, the measuring site and their coupling.  We further have vector currents coming from the unit vectors and the magnetic field. To arrive at the last form for the currents we used the detailed form of the 2 by 2 matrix $h(\vec{q})$ to invert it and evaluate at zero momentum; while also evaluating $A_{\sigma'}(0)=3(1+\gamma)$ . The partition function then becomes:
\begin{align}
    Z_{\text{text}}[\tilde{J}_z]= \int \mathcal{D} \vec{n}_1 \int \mathcal{D} \vec{n}_2 \int \mathcal{D} \vec{\mu}\exp{-S_{\text{text}}^{\mu}+\tilde{J}_z n_2^z+\dfrac{1}{2}\beta^2Nh^2\sum_{\sigma,\sigma'}h^{-1}_{\sigma\sigma'}(0)}.
\end{align}
 As noted before this action is again quadratic in the  $\vec{\mu}_i$ fields and can be then integrated out. We will arrive at an expression concerning only the degrees of freedom from the snowflake and the measuring site, in mathematical terms we get:
 \begin{align}
      Z_{\text{text}}[\tilde{J}_z]=\int \mathcal{D} \vec{n}_1 \int \mathcal{D} \vec{n}_2 \exp{\dfrac{1}{2}J_s^\nu D^{-1}_{s,s'}J_{s'}^\nu+\tilde{J}_z n_2^z+\frac{N \beta^2h^2}{18 \beta  (\gamma +1)^2+\rho }},
 \end{align}
 \begin{align}
     D^{-1}=\dfrac{1}{D_1/3-D_{12}^2}\begin{pmatrix}
         1/3 && -D_{12}\\
         -D_{12} && D_1
     \end{pmatrix},\qquad D_{1}=\dfrac{1}{N} \sum_{\vec{q},\sigma,\sigma'} A_{\sigma}(\vec{q})h^{-1}_{\sigma\sigma'}(\vec{q})A_{\sigma'} (-\vec{q}), \quad D_{12}=\dfrac{1}{N}\sum_{\vec{q},\sigma} A_{\sigma}(\vec{q}) e^{i\vec{q}\cdot (\vec{R}_2+\vec{\tau}_{\sigma_2})}h^{-1}_{\sigma\sigma_2}(\vec{q}).
 \end{align}
 The matrix $D^{-1}_{s,s'}$ is the inverse of the matrix describing the quadratic form in $\mu_i^\nu$ . We used the fact that $D_{22}=\dfrac{1}{N}\sum_{\vec{q}}h^{-1}_{\sigma_2\sigma_2}(\vec{q})=1/3$ since it is equal to $\expval{\vec{\phi}_2^2}/3$,  by the previous definitions. It is easy now to expand the vector currents in terms of the unit vectors describing the snowflake degree of freedom and the measuring site so as to obtain the effective action :
 \begin{align}
     S_{\text{eff}}[\tilde{J}_z]= -\left( \dfrac{\vec{J}_1^2/2}{D_1-3 D_{12}^2}-\dfrac{3 D_{12}\vec{J}_1\cdot\vec{J}_2}{D_1-3D_{12}^2} +\dfrac{3D_1\vec{J}_2^2/2}{D_1-3 D_{12}^2}\right)-\tilde{J}_z n_2^z-\frac{N \beta^2h^2}{18 \beta  (\gamma +1)^2+\rho }.
 \end{align}
 Before proceeding  with the calculation of the spin texture it is worth noting that at this level of approximation we can obtain also the orphan spin magnetic moment and compare with the previous two methods. The difference in the approximation between the last section and this one is essentially on approximating $\mathcal{P}_OC\mathcal{P}_O$ to be given by the 2 by 2 matrix $D$. This is true if the internal structure of the orphan snowflake is not important. We specialize to the low temperature regime which implies:
 \begin{align}
     -\dfrac{\vec{J}_1^2/2}{D_1-3 D_{12}^2}\approx-\dfrac{1}{2}\dfrac{1}{D_1-3 D_{12}^2}\left(-\frac{i \vec{h}\ }{3  (\gamma +1) }+i\vec{n}_1\right)^2\approx-\dfrac{1}{2 D_1}\left(-\frac{i \vec{h}\ }{3  (\gamma +1) }+i\vec{n}_1\right)^2,
 \end{align}
  where we approximate the position of the measurement to be far enough from the orphan so that $D_{12}=\dfrac{1}{3}\expval{\vec{\phi}^1_\gamma\cdot \vec{\phi}_2}\approx 0$. This is valid if we are interested only in the magnetization, since we are averaging over all the measuring sites; which effectively reduces the problem to just looking at the snowflake orphan spin. Thus this approximation will not be valid for the local spin texture where the measuring site is fixed to a definite position.  Let us now use this approximation to simplify the effective action in terms of the unit vectors  meaning:
 \begin{align}
      S_{\text{eff}}[\tilde{J}_z]\approx -\dfrac{1}{D_1 \ 3  (\gamma +1)}\vec{h}\cdot \vec{n}_1+ \dfrac{1}{2D_1}+\frac{h^2}{18 D_1  (\gamma +1)^2 }+\dfrac{3 D_{12}\vec{J}_1\cdot\vec{J}_2}{D_1-3D_{12}^2} -\dfrac{3D_1\vec{J}_2^2/2}{D_1-3 D_{12}^2}-\tilde{J}_z n_2^z-\frac{N \beta h^2}{18  (\gamma +1)^2 }.
 \end{align}
To proceed further we need to calculate $D_1=\dfrac{1}{3}\expval{\vec{\phi}^1_\gamma\cdot \vec{\phi}^1_\gamma}$ . We do this by focusing on the undiluted spin liquid action and assuming $\beta\rightarrow \infty, h\rightarrow 0$ . This limits can then be applied to the undiluted correlator to decoupled the clusters and give rise to a simpler description:
 \begin{align}
     \expval{\vec{\phi}^O_\gamma\cdot \vec{\phi}^O_\gamma}&=\dfrac{1}{Z_{\text{ud}}}\int \mathcal{D} \vec{\phi} \ (\vec{\phi}^O_\gamma)^2 \  e^{-\frac{\beta}{2}\sum_{\text{hex}} \vec{\phi}_\gamma^2+\beta \vec{h} \sum_i \vec{\phi}_i}=\sum_\nu \dfrac{1}{Z_{\text{ud}}}\dfrac{\delta^2}{\delta \tilde{J}_\nu \delta \tilde{J}_\nu}\int \mathcal{D} \vec{\phi} \  e^{-\frac{\beta}{2}\sum_{\text{hex}} \vec{\phi}_\gamma^2+\beta \vec{h} \sum_i \vec{\phi}_i+\tilde{J}\vec{\phi}^O_\gamma}\eval_{\tilde{J}=0}\\
     & \approx \sum_\nu \dfrac{1}{Z_{\text{ud}}}\dfrac{\delta^2}{\delta \tilde{J}_\nu \delta \tilde{J}_\nu}\int \mathcal{D} \vec{\phi} \  e^{-\frac{\beta}{2}\sum_{\text{hex}-O} \vec{\phi}_\gamma^2-\frac{\beta}{2}(\vec{\phi}^O_\gamma)^2+\tilde{J}\vec{\phi}^O_\gamma}\eval_{\tilde{J}=0} \approx\sum_\nu \dfrac{1}{Z_{\text{O}}}\dfrac{\delta^2}{\delta \tilde{J}_\nu \delta \tilde{J}_\nu}\int_{\text{O}} \mathcal{D} \vec{\phi} \  e^{-\frac{\beta}{2}(\vec{\phi}^O_\gamma)^2+\tilde{J}\vec{\phi}^O_\gamma}\eval_{\tilde{J}=0}\\
     &=\sum_\nu \dfrac{1}{Z_{\text{O}}}\dfrac{\delta^2}{\delta \tilde{J}_\nu \delta \tilde{J}_\nu}\int_{\text{O}} \mathcal{D} \vec{\phi} \int \mathcal{D} \vec{L}_\gamma \delta(\vec{L}_\gamma-\vec{\phi}^O_\gamma) \  e^{-\frac{\beta}{2}(\vec{L}_\gamma)^2+\tilde{J}\vec{L}_\gamma}\eval_{\tilde{J}=0} \approx \sum_\nu \dfrac{1}{Z_{\text{O}}}\dfrac{\delta^2}{\delta \tilde{J}_\nu \delta \tilde{J}_\nu} \int \mathcal{D} \vec{L}_\gamma  \  e^{-\frac{\beta}{2}(\vec{L}_\gamma)^2+\tilde{J}\vec{L}_\gamma +\ln{F}}\eval_{\tilde{J}=0},
 \end{align}
 where we have introduced a generating field $\tilde{J}_\nu$ and approximated the snowflake constraint vectors to be independent so the contributions from the numerator cancel the $Z_{\text{ud}}$ denominator and leave only the orphan snowflake partition function. We have then used a Dirac Delta identity to express the integral over all the spins in the cluster in terms of the constraint vector only. The last approximation considers that $\int \mathcal{D} \vec{\phi}\ \delta(\vec{L}_\gamma-\vec{\phi}^O_\gamma)$ is a constant since it basically counts the number of ways the spins can be arranged to produce a constraint vector of size $ \vec{L}_\gamma$, which is independent of $\tilde{J}$ and temperature, for small enough temperatures. Since we only care about the functional derivative this constant will cancel with the $Z_{O}$ in the denominator. We can now solve the Gaussian integral in  $ \vec{L}_\gamma$  to obtain:
 \begin{align}
     D_1=\dfrac{1}{3}\expval{\vec{\phi}^1_\gamma\cdot \vec{\phi}^1_\gamma}\approx \dfrac{1}{3}\sum_\nu \dfrac{\delta^2}{\delta \tilde{J}_\nu \delta \tilde{J}_\nu} \  e^{\frac{1}{2 \beta}\tilde{J}^\alpha \tilde{J}^\alpha}\eval_{\tilde{J}=0} =\dfrac{1}{\beta}\ .
 \end{align}
We see that under the previous approximations the behaviour of the orphan snowflake correlator is simply proportional to the temperature. Neglecting constant energy shifts, the effective action becomes:
 \begin{align}
     S_{\text{eff}}[\tilde{J}_z]\approx -\beta\vec{h}\cdot (\alpha\vec{n}_1)+ +\dfrac{3 D_{12}\vec{J}_1\cdot\vec{J}_2}{D_1-3D_{12}^2} -\dfrac{3D_1\vec{J}_2^2/2}{D_1-3 D_{12}^2}-\tilde{J}_z n_2^z\ .
 \end{align}
From this it follows that if one is interested only in the orphan cluster, the relevant contribution to the magnetization is the first term, which gives the expected irrational magnetic moment of:
 \begin{align}
     \alpha=\dfrac{1}{3(1+\gamma)}.
 \end{align}

 This is the result we got from the naive Lagrange multiplier formula without considering the full hybrid field theory. Let us now focus on the spin texture , which we can compute now by writing out explicitly the action in terms of the unit vectors:
 \begin{align}
         S_{\text{eff}}[\tilde{J}_z]=\dfrac{3/2}{1-3 \beta D_{12}^2}\vec{n}_2^2+\beta\left(-\alpha+\dfrac{3\alpha^2 D_{12}/2}{1-3\beta D_{12}^2}\right)\vec{n}_1\cdot \vec{h}-\beta \dfrac{3 D_{12}}{1-3\beta D_{12}}\vec{n}_1\cdot\vec{n}_2+\vec{n}_2\cdot \left( -\tilde{J}_z\hat{e}_z-\dfrac{3\alpha^2/2}{1-3 \beta D_{12}^2}\vec{h}\right),
 \end{align}
  where we have neglected constants that do not depend on the unit vectors, since the expectation value has a partition function in the denominator and will cancel these terms. Let us assume now that the spin at the measuring position is a soft spin as a first approximation, mathematically this means the measure has no delta function now, so that the partition function becomes:
 \begin{align}
     &Z_{\text{text}}[\tilde{J}_z]\approx \int \mathcal{D} \vec{n}_1 e^{-S_1'[\vec{n}_1]}\int \mathcal{D} \vec{\phi}_2 \exp{-\dfrac{1}{2}W\phi_2^\alpha \phi_2^\alpha+K_2^\alpha \phi_2^\alpha}\\
     &W=\dfrac{3}{1-3\beta D_{12}^2},\quad \vec{K}_2=\tilde{J}_z\hat{e}_z+\dfrac{3}{2}\dfrac{\alpha^2}{1-3\beta D_{12}^2}\vec{h}+\dfrac{3\beta D_{12}}{1-3\beta D_{12}^2}\vec{n}_1, \quad S_1'[\vec{n}_1]=\beta\left(-\alpha+\dfrac{3\alpha^2 D_{12}/2}{1-3\beta D_{12}^2}\right)\vec{n}_1\cdot \vec{h},
 \end{align}
 where we observe that the action separates into a part independent of the soft spin $S_1'$ and a Gaussian part  in terms of $\phi_2$. When integrated out leads to an effective action in terms of just the orphan spin vector $\vec{n}_1$:
 \begin{align}
     Z_{\text{text}}[\tilde{J}_z]=\int \mathcal{D} \vec{n}_1 e^{-S_{\text{text}}[\vec{n}_1,\tilde{J}_z]}=\int \mathcal{D} \vec{n}_1 \exp{-\beta\left(-\alpha+\dfrac{3\alpha^2 D_{12}/2}{1-3\beta D_{12}^2}\right)\vec{n}_1\cdot \vec{h}+\dfrac{1}{2 W} \vec{K}_2^2}\ .
 \end{align}
Remarkably the action is again that of a free spin in a modified magnetic field and a tunable magnetic moment. Because of the fixed measurement position we also have a dependence on the position coming from $D_{12}$ the correlator of the undiluted spin liquid. The effective action can be rewritten in a simpler way :
 \begin{align}
   &S_{\text{text}}[\vec{n}_1,\tilde{J}_z]=\beta\left(-\alpha+\dfrac{\alpha^2}{2} D_{12}W\right)n_1^z h-\dfrac{1}{2 }\left(\tilde{J}_z^2+\alpha^2  h\tilde{J}_z+2 n_1^z\beta D_{12}\tilde{J}_z+W\beta \alpha^2 D_{12}n_1^z h\right)\\
   &S_{\text{text}}[\vec{n}_1,\tilde{J}_z]=-\beta(\alpha n_1^z)h-\dfrac{1}{2}\tilde{J}_z^2-\tilde{J}_z\left(\dfrac{1}{2}\alpha^2h +n_1^z\beta D_{12}\right)\ ,
 \end{align}
where we again have neglected constants in energy which don't depend on the current or orphan spin vector and assumed the magnetic field to point in the $z$ direction. Finally we want to calculate the first derivative with respect to the current and to evaluate at zero current so the quadratic part will not play a role, we are thus left with:
 \begin{align}
     &\expval{S^z_2}=\dfrac{1}{Z_{\text{text}}[\tilde{J}_z]}\dfrac{\delta Z_{\text{text}}[\tilde{J}_z]}{\delta \tilde{J}_z}\eval_{\tilde{J}_z=0}=\dfrac{1}{Z_{\text{text}}[0]}\int \mathcal{D} \vec{n}_1 e^{-S_{\text{text}}[\vec{n}_1,0]}\left(\dfrac{1}{2}\alpha^2h +n_1^z\beta D_{12}\right)\\
     &\expval{S^z_2}=\dfrac{1}{Z_{\text{text}}[0]}\int \mathcal{D} \vec{n}_1 e^{\beta(\alpha n_1^z)h}\left(\dfrac{1}{2}\alpha^2h +n_1^z\beta D_{12}\right)=\dfrac{1}{2}\alpha^2 h +\beta  D_{12}\ L \left( \alpha \beta h \right),
 \end{align}
 The final result is simple and depends linearly on the magnetic field as well as through the Langevin function with the undiluted spin liquid correlator evaluated at the measurement site:
 \begin{align}
     \expval{S^z_2}=\dfrac{1}{2}\alpha^2 h +\dfrac{1}{3}\beta \expval{\vec{\phi}^1_\gamma\cdot \vec{\phi}_2} \ L \left( \alpha \beta h \right)
 \end{align}

\subsection{Scaling of the charge-spin correlation function at the higher-rank point }

We see now that the position dependence comes entirely from the charge-spin correlation function, which we can calculate for the special point $\gamma=1/2$ by expanding it's momentum space expression near the $K,K'$ point , which is where the gapless point happens, first we expand the interaction matrix near $\vec{q}_0=\vec{K}$ for some small momentum $\Delta \vec{q}=\vec{q}-\vec{q}_0\equiv\vec{k}$ up to fourth order in $1/a$ i.e. inverse lattice length:
\begin{align}
    h(\vec{q})&=h(\vec{q}_0+\Delta \vec{q})\approx(\rho\sigma_0+\dfrac{g}{2}\beta \vec{\abs{k}}^4)\sigma_0+2g\beta (k_x^3k_y-k_y^3k_x)\sigma_y+\dfrac{g}{2}\beta(\vec{\abs{k}}^4-8k_x^2k_y^2)\sigma_x\\
    &=\rho\sigma_0+\dfrac{g}{2}\beta \begin{pmatrix}
        \vec{\abs{k}}^4 && (k_x-ik_y)^4\\
        (k_x+ik_y)^4 && \vec{\abs{k}}^4
    \end{pmatrix},
\end{align}
 where we defined the microscopic constant $g=9/32$, we have also used the set of coordinates $k_x,k_y$ such that $k_x\hat{e}_x+k_y\hat{e}_y=\vec{k}=k'_x\vec{b}_1+k'_y\vec{b}_2$. Where $\vec{b}_{1,2}$ are the reciprocal lattice vectors with corresponding lattice vectors $\vec{a}_1=a\hat{e}_x,\vec{a}_2=a\hat{e}_x/2+\sqrt{3}a\hat{e}_y/2$ now the inverse can be approximated to be:
\begin{align}
    h^{-1}(\vec{k})\approx \dfrac{1}{\left(\rho+\dfrac{g}{2}\beta\vec{\abs{k}}^4\right)^2-\left(\dfrac{g}{2}\beta\right)^2\vec{\abs{k}}^8} \ \left(\rho\sigma_0+\dfrac{g}{2}\beta \begin{pmatrix}
        \vec{\abs{k}}^4 && -(k_x-ik_y)^4\\
        -(k_x+ik_y)^4 && \vec{\abs{k}}^4
    \end{pmatrix}\right).
\end{align}
We need to be consistent with the order of approximation so that both numerator and denominator have the same order which implies then:
\begin{align}
    h^{-1}(\vec{k})\approx \dfrac{1}{\rho^2+g\rho\beta\vec{\abs{k}}^4} \ \left(\rho\sigma_0+\dfrac{g}{2}\beta \begin{pmatrix}
        \vec{\abs{k}}^4 && -(k_x-ik_y)^4\\
        -(k_x+ik_y)^4 && \vec{\abs{k}}^4
    \end{pmatrix}\right).
\end{align}
Already at this level we see that the inverse interaction matrix has no usual ${k}^2$ dependence and has instead to lowest order behaves as ${k}^4$. Next we expand $A_{\sigma}(\vec{q})$ around the same $K$ point to the third order to obtain:
\begin{align}
    A_K(\vec{q})=A(\vec{q}_0+\Delta \vec{q})\approx\dfrac{3}{8}\begin{pmatrix}
        (k_x+ik_y)^2\\
         (k_x-ik_y)^2
    \end{pmatrix},
\end{align}
analogously expanding around the other $K'$ point we have:
\begin{align}
    A_{K'}(\vec{q})=A(\vec{q}_0+\Delta \vec{q})\approx\dfrac{3}{8}\begin{pmatrix}
        (k_x-ik_y)^2\\
         (k_x+ik_y)^2
    \end{pmatrix}.
\end{align}
It is clear now that the contributions to lowest order are purely quartic in the momentum. We come now to the correlation function which using the previous approximations can be cast into an integral form in the continuum and to second order in the numerator gives:
\begin{align}
    D_{12}&=\dfrac{1}{N}\sum_{\vec{q},\sigma} A_{\sigma}(\vec{q}) e^{i\vec{q}\cdot (\vec{R}_2+\vec{\tau}_{\sigma_2})}h^{-1}_{\sigma\sigma_2}(\vec{q})\notag \\
    &\approx e^{i\vec{K}\vec{r}_2} \dfrac{3}{8 (2\pi)^2}\int^{\Lambda} \dd^2 \vec{k}\ \dfrac{(k_x+i\eta(\vec{r}_2) k_y)^2}{\rho+g\beta  \vec{\abs{k}}^4}e^{i\vec{k}\cdot \vec{r}_{12}}+e^{i\vec{K}'\vec{r}_2} \dfrac{3}{8 (2\pi)^2}\int^{\Lambda} \dd^2 \vec{k}\ \dfrac{(k_x-i\eta(\vec{r}_2) k_y)^2}{\rho+g\beta  \vec{\abs{k}}^4}e^{i\vec{k}\cdot \vec{r}_{12}},
\end{align}
  where we introduced the variable
  $\eta(\vec{r}_1)=\pm 1$ depending on which sub-lattice the position of the spin sis measured at. Before approximating this integral let us note we introduced a cut-off momentum $\Lambda$ which takes into account when the dispersion expansion near the K point is no longer valid. We can now examine the scaling with temperature as for other classical spin liquids by a change of variables $k\rightarrow k T^{1/4}$ which makes the denominator not depend on temperature and we define $\vec{x}=\vec{r}_{12}T^{1/4}$ and $k_x+i \eta_2 k_y =ke^{i\eta_2 \theta} $, with $K=-K'$ so we obtain:
\begin{align}
        \expval{\vec{\phi}^1_\gamma\cdot \vec{\phi}_2} \approx  e^{i\vec{K}\vec{r}_2} \dfrac{3 T}{8 (2\pi)^2}\int^{\Lambda/T^{1/4}} \dd^2 \vec{k} \ \ \dfrac{k^2 e^{ 2 i\eta_2 \theta} }{\rho+g  \vec{\abs{k}}^4}e^{i\vec{k}\cdot \vec{x}}+ e^{-i\vec{K}\vec{r}_2} \dfrac{3 T}{8 (2\pi)^2}\int^{\Lambda/T^{1/4}} \dd^2 \vec{k} \ \ \dfrac{k^2e^{- 2 i\eta_2 \theta} }{\rho+g  \vec{\abs{k}}^4}e^{i\vec{k}\cdot \vec{x}}.
\end{align}
We observe now an interesting scaling behaviour of the form:
\begin{align}
     \expval{\vec{\phi}_\gamma (\vec{r}_1)\cdot \vec{\phi}(\vec{r}_2)} \propto T F_1((\vec{r}_{1}-\vec{r}_{2}) T^{1/4}).
\end{align}
This result is already different from a typical U(1) classical spin liquid, where the low energy theory result for the scaling of the charge-spin correlator always implies a proportionality constant like $T^{3/2}$, and a function that depends only on the magnitude of the distance and with a temperature dependent factor in the argument of $T^{1/2}$; crucially, here we observe  a nontrivial angular dependence.\\
Let us now look at the scaling behaviour for large and small $\vec{x}$. Since we can fix the x,y axis and measuring the vector $\vec{k}$ at the polar angle $\theta$ and the vector $\vec{x}$ at the polar angle $\phi$ we obtain:
\begin{align}
    \expval{\vec{\phi}^1_\gamma\cdot \vec{\phi}_2} = \dfrac{3 T}{8 (2\pi)^2} \int_{0}^{\infty}\int_0^{2\pi}\text{k}  \dd \text{k}  \dd \theta  \ \ \frac{\text{k}^2 }{\rho+g  \text{k}^4}(e^{2 i\eta_2 \theta+i\vec{K}\cdot \vec{r}_2}+e^{-2 i\eta_2 \theta-i\vec{K}\cdot \vec{r}_2})e^{i\text{k} x\cos{(\theta-\phi)}}.
\end{align}
In this way we have collected two integrals into one so as to facilitate the next steps for solving them. Let us use now the Jacobi-Anger identity to rewrite the exponential in terms of Bessel functions:
\begin{align}
    \int_{0}^{\infty}\int_0^{2\pi}\text{k}  \dd \text{k}  \dd \theta  \ \ \frac{\text{k}^2 e^{\pm 2 i\eta_2 \theta}}{\rho+g  \text{k}^4}e^{i\text{k} x\cos{(\theta-\phi)}}=\sum_{n=-\infty}^{\infty} \int_{0}^{\infty}\int_0^{2\pi}\text{k}  \dd \text{k}  \dd \theta  \ \ \frac{\text{k}^2}{\rho+g  \text{k}^4} i^n J_{n}(\text{k} x)e^{in(\theta-\phi)\pm 2 i\eta_2 \theta},
\end{align}
the angular integral is always zero except for $n=\mp 2 \eta_2 $ for which it gives $2 \pi $ , there is now only a radial integral remaining:
\begin{align}
    \expval{\vec{\phi}^1_\gamma\cdot \vec{\phi}_2} = - \dfrac{3 T}{8\pi} \int_{0}^{\infty}\text{k}  \dd \text{k} \ \frac{\text{k}^2 }{\rho+g  \text{k}^4}  J_{2}(\text{k} x) \dfrac{1}{2}(e^{i 2 \eta_2 \phi +i\vec{K}\cdot \vec{r}_2}+e^{-i 2 \eta_2 \phi -i\vec{K}\cdot \vec{r}_2})=- \dfrac{3 T}{8\pi} \cos{(\vec{K}\cdot \vec{r}_2+2\eta_2 \phi)} \int_{0}^{\infty} \dd \text{k} \ \frac{\text{k}^2 }{\rho+g  \text{k}^4}  J_{2}(\text{k} x),
\end{align}
where we used the Bessel function identity $J_{-m}(x)=(-1)^mJ_{m}(x)$. The last integral is actually the well-known Hankel transform of order $\nu=2$ of the function $\text{k}^2/(\rho+g \text{k}^4)$ which can be expressed in terms of the Meijer G-function:
 \begin{align}
 \expval{\vec{\phi}^1_\gamma\cdot \vec{\phi}_2} = - \dfrac{24 T}{\pi} \cos{(\vec{K}\cdot \vec{r}_2+2\eta_2 \phi)}\dfrac{1 }{\rho \abs{\vec{x}}^4}  \displaystyle G^{\, 03}_{40}\!\left(\left.{\begin{matrix}-1,\  -1/2,\ 0,\ 1/2 \\ - \end{matrix}}\;\right|\,\dfrac{256 g}{\abs{\vec{x}}^4 \rho}\right).
 \end{align}
We observe the radial decay is contained in the last part while the cosine only represents a modulation depending on the UV momentum $\vec{K}$ from the lattice and a polar dependence which changes sign from one sublattice to the other. We can further expand around $x\ll 1$ , to lowest order in $x$ we obtain:
\begin{align}
    F_1(\vec{x}) \approx \cos{(\vec{K}\cdot \vec{r}_2+2\eta_2 \phi)}\left(\frac{\pi}{ g}-\frac{\pi^2  \abs{\vec{x}}^2 \sqrt{\rho}}{16g^{3/2}}+\mathcal{O}(x^4)\right),
\end{align}
while for $x\gg 1$ we obtain to lowest order:
\begin{align}
    F_1(\vec{x}) \approx \dfrac{\cos{(\vec{K}\cdot \vec{r}_2+2\eta_2 \phi)}}{\rho \t{x}^4}\displaystyle G^{\, 03}_{40}\!\left(\left.{\begin{matrix}-1,\  -1/2,\ 0,\ 1/2 \\ - \end{matrix}}\;\right|\,0\right)
\end{align}

 We can alternatively coarse grain from the start and take the rank-2 U(1) gauge theory and examine the correlator from this framework. Doing this requires considering the simplest action which reproduces to the ground state constraint in terms of a traceless symmetric field $m_{\mu\nu}(x,y)$ with the constraint being $\partial_\mu\partial_\nu m_{\mu\nu}=0$, this means we postulate a partition function of the form:
\begin{align}
    Z_{m}=\int \mathcal{D}m e^{-S_m}= \int \mathcal{D}m \exp{-\int \dd^2 \vec{r}\left(\dfrac{\lambda}{2}m_{\mu\nu}m_{\mu\nu}+\Delta(\partial_\mu\partial_\nu m_{\mu\nu})^2 \right)}
\end{align}

\section{Interaction between orphan spins}

In this section we want to explore what is the effect of having two different orphan clusters separated a fixed distance. We will calculate the effective spin exchange interaction, as well as the spin-spin correlation function between the two orphan spins. To do this we will denote by $1,2$ the two different orphans at positions $\vec{R}_1=0,\vec{R}_2$. We further use the approximation that two orphan spins, far away from each other for sufficiently large distances can be consider to modify the partition function by just fixing the constraint vector to be a fixed length vector. This is very similar to the previous spin texture calculation, except that now we apply this to both clusters. In other words we can express the partition function of the diluted spin liquid as:
\begin{align}
    Z_{\text{int}}&= \int \mathcal{D} \vec{\phi}\int \mathcal{D} \vec{n}_1 \ \delta(\vec{\phi}^1_\gamma-\vec{n}_1 )\int \mathcal{D} \vec{n}_2  \ \delta(\vec{\phi}^2_\gamma-\vec{n}_2 ) \ e^{-S_{\text{ud}}[\vec{\phi},\vec{h}]} \\
    &= \int \mathcal{D} \vec{\phi}\int \mathcal{D} \vec{n}_1 \int \mathcal{D} \vec{n}_2 \int \mathcal{D} \vec{\mu}_1 \int \mathcal{D} \vec{\mu}_2 \exp{i(\vec{\phi}^1_\gamma-\vec{n}_1 )\cdot \vec{\mu}_1+i(\vec{\phi}^2_\gamma-\vec{n}_2 )\cdot \vec{\mu}_2-S_{\text{ud}}}.
\end{align}
Just as in the last section we rewrite the action in terms of the Fourier basis where we assume the orphan spin located at the unit cell $\vec{R}_1$ to be at the origin of the lattice. We can then express the Lagrange multiplier part of the action just as we did before for the spin texture:
\begin{align}
    i\vec{\phi}^1_\gamma \cdot \vec{\mu}_1 &= \dfrac{i}{\sqrt{N}}\left(\sum_{\vec{q}}\sum_{i \in \mathrm{hex}} \vec{\Phi}_{\vec{q}}^{\sigma(i)}e^{-i\vec{q}\cdot \vec{r}(i)}+\gamma \sum_{i \in\langle\mathrm{hex}\rangle} \vec{\Phi}_{\vec{q}}^{\sigma(i)}e^{-i\vec{q}\cdot \vec{r}(i)}\right) \cdot \vec{\mu}_1
    =\dfrac{i}{\sqrt{N}}\left( \sum_{\vec{q},\sigma} \sum_{\vec{r}_h \in \sigma } \vec{\Phi}_{\vec{q}}^{\sigma}e^{-i\vec{q}\cdot \vec{r}_h }+ \gamma \sum_{\vec{r}_{\expval{h}} \in \sigma } \vec{\Phi}_{\vec{q}}^{\sigma}e^{-i\vec{q}\cdot \vec{r}_{\expval{h}} } \right) \cdot \vec{\mu}_1\\
    &=  \sum_{\vec{q},\sigma} \vec{J}^{1}_{\vec{q}\sigma} \vec{\Phi}_{\vec{q}}^{\sigma}, \qquad  \vec{J}^{1}_{\vec{q}\sigma} = \dfrac{i}{\sqrt{N}} \left(\sum_{\vec{r}_h \in \sigma } e^{-i\vec{q}\cdot \vec{r}_h }+ \gamma \sum_{\vec{r}_{\expval{h}} \in \sigma } e^{-i\vec{q}\cdot \vec{r}_{\expval{h}} } \right)\vec{\mu}_1,
\end{align}
where we have defined the vector current, which couples to the Lagrange multiplier $\vec{\mu}_1$ for the orphan cluster at the origin. Similarly we can now calculate the contribution to the action which couples to $\vec{\mu}_2$ . It is worth mentioning that the previous factor , inside parenthesis defining $ \vec{J}^{2}_{\vec{q}\sigma} $ , is the same that appeared for the spin texture , which we denoted as $A_{\sigma}(\vec{q})$ . Rewriting the equation for $ \vec{J}^{2}_{\vec{q}\sigma} $in terms of this quantity we have :
\begin{align}
    i\vec{\phi}^2_\gamma \cdot \vec{\mu}_2 &= \dfrac{i}{\sqrt{N}}\left(\sum_{\vec{q}}\sum_{i \in \mathrm{hex}} \vec{\Phi}_{\vec{q}}^{\sigma(i)}e^{-i\vec{q}\cdot \vec{r}_2(i)}+\gamma \sum_{i \in\langle\mathrm{hex}\rangle} \vec{\Phi}_{\vec{q}}^{\sigma(i)}e^{-i\vec{q}\cdot \vec{r}_2(i)}\right) \cdot \vec{\mu}_2
    =\dfrac{i}{\sqrt{N}}\sum_{\vec{q},\sigma} e^{-i\vec{q}\cdot \vec{R}_2}A_{\sigma}(\vec{q})\vec{\Phi}_{\vec{q}}^{\sigma} \cdot \vec{\mu}_2\\
    &=  \sum_{\vec{q},\sigma} \vec{J}^{2}_{\vec{q}\sigma} \vec{\Phi}_{\vec{q}}^{\sigma}, \qquad  \vec{J}^{2}_{\vec{q}\sigma} = \dfrac{i}{\sqrt{N}} e^{-i\vec{q}\cdot \vec{R}_2}A_{\sigma}(\vec{q})\vec{\mu}_2,
\end{align}
We note here how similar the calculation is to the spin texture, the difference will come when calculating the correlator. It is clear now that  grouping this terms together into a single current vector we obtain:
\begin{align}
    S_{\text{int}}=-i(\vec{\phi}^1_\gamma-\vec{n}_1 )\cdot \vec{\mu}_1-i(\vec{\phi}^2_\gamma-\vec{n}_2 )\cdot \vec{\mu}_2+S_{\text{ud}}=\frac{1}{2}  \sum_{\vec{q},\sigma,\sigma'}\vec{\Phi}^{\sigma}_{-\vec{q}}\cdot h_{\sigma \sigma'}(\vec{q}) \vec{\Phi}^{\sigma'}_{\vec{q}}-\sum_{\vec{q},\sigma} \vec{J}_{\vec{q}\sigma} \vec{\Phi}_{\vec{q}}^{\sigma}+i(\vec{n}_1\cdot \vec{\mu}_1+\vec{n}_2\cdot \vec{\mu}_2),
\end{align}
where the effective action has now given rise to a total vector current coupling to the $\phi$ field just as before. The action is at this point quadratic, which means we can integrate out the $\vec{\phi}$ field by performing the Gaussian integral so as to give:
\begin{align}
    Z_{\text{int}}= \int \mathcal{D} \vec{n}_1 \int \mathcal{D} \vec{n}_2 \int \mathcal{D} \vec{\mu}_1 \int \mathcal{D} \vec{\mu}_2 \exp{\dfrac{1}{2}\sum_{\vec{q},\sigma,\sigma'}\vec{J}_{\vec{q}\sigma}h^{-1}_{\sigma\sigma'}(\vec{q})\vec{J}_{-\vec{q}\sigma'}-i(\vec{n}_1\cdot \vec{\mu}_1+\vec{n}_2\cdot \vec{\mu}_2)},
\end{align}
where now we are left only with the Lagrange multiplier fields, as well as the unit length vectors $\vec{n}_1$ and $\vec{n}_2$ . If we expand the action in terms of the previous definition for the total vector current we obtain:
\begin{align}
    S_{\text{text}}^{\text{eff}}&=-\dfrac{1}{2 N}\sum_{\vec{q},\sigma,\sigma'}\left(i A_{\sigma}(\vec{q})\vec{\mu}_1+i A_{\sigma}(\vec{q})e^{-i\vec{q}\cdot \vec{R}_2} \vec{\mu}_2\right)h^{-1}_{\sigma\sigma'}(\vec{q})\left(i A_{\sigma'}(-\vec{q})\vec{\mu}_1+i e^{i\vec{q}\cdot \vec{R}_2} A_{\sigma'}(-\vec{q})\vec{\mu}_2 \right)\notag\\
    &+i(\vec{n}_1\cdot \vec{\mu}_1+\vec{n}_2\cdot \vec{\mu}_2),
\end{align}
 where we assumed no magnetic field needs to be present to have a response between the two orphan spins. Let us rewrite the effective action in a matrix form just as in the previous section, so as to integrate the $\mu^\alpha_i$ field:
\begin{align}
    S_{\text{text}}^{\mu}&=\dfrac{1}{2 N}\begin{pmatrix}
        \mu_1^\nu  && \mu_2^\nu
    \end{pmatrix}
    \begin{pmatrix}
        \sum_{\vec{q},\sigma,\sigma'} A_{\sigma}(\vec{q})h^{-1}_{\sigma\sigma'}(\vec{q})A_{\sigma'} (-\vec{q}) && \sum_{\vec{q},\sigma} A_{\sigma}(\vec{q}) e^{i\vec{q}\cdot \vec{R}_2
        }h^{-1}_{\sigma\sigma'}(\vec{q})A_{\sigma'}(-\vec{q}) \\
        \sum_{\vec{q},\sigma} A_{\sigma}(\vec{q}) e^{-i\vec{q}\cdot \vec{R}_2
        }h^{-1}_{\sigma\sigma'}(\vec{q})A_{\sigma'}(-\vec{q})  &&  \sum_{\vec{q},\sigma,\sigma'} A_{\sigma}(\vec{q})h^{-1}_{\sigma\sigma'}(\vec{q})A_{\sigma'} (-\vec{q}) 
    \end{pmatrix}\begin{pmatrix}
        \mu_1^\nu  \\ \mu_2^\nu
    \end{pmatrix}+i\vec{n}_1\cdot \vec{\mu}_1+i\vec{n}_2\cdot \vec{\mu}_2 .
\end{align}
We recognize the structure of the matrix as the one we had for the spin texture except that now the diagonal has the same function. Since both Lagrange multipliers relate to far away orphan spins, no distinction at this level of approximation can arise.  Once again we are left with a Gaussian integral in the  $\vec{\mu}_i$ fields, we integrate them out to obtain an effective action in terms of the $\vec{n}_i$ vectors:
 \begin{align}
      Z_{\text{text}}[\tilde{J}_z]=\int \mathcal{D} \vec{n}_1 \int \mathcal{D} \vec{n}_2 \exp{-\dfrac{1}{2}n_s^\nu G^{-1}_{s,s'}n_{s'}^\nu},
 \end{align}
 where the matrix $G^{-1}_{s,s'}$ is the inverse of the matrix describing the quadratic form in $\mu_i^\nu$ given by:
 \begin{align}
     G^{-1}=\dfrac{1}{D_1^2-G_{12}^2}\begin{pmatrix}
         D_1 && -G_{12}\\
         -G_{12} && D_1
     \end{pmatrix},\qquad D_{1}=\dfrac{1}{N} \sum_{\vec{q},\sigma,\sigma'} A_{\sigma}(\vec{q})h^{-1}_{\sigma\sigma'}(\vec{q})A_{\sigma'} (-\vec{q}), \quad G_{12}=\dfrac{1}{N} \sum_{\vec{q},\sigma,\sigma'}e^{i\vec{q}\cdot \vec{R}_2
        } A_{\sigma}(\vec{q}) h^{-1}_{\sigma\sigma'}(\vec{q})A_{\sigma'}(-\vec{q}).
 \end{align}
It is clear here how $D_1$ appears for both the spin texture and the interaction between orphan spins while $G_{12}$ , the charge-charge correlation function, has an extra factor of $A_{\sigma}(\vec{q})$ arising from the fact that each orphan spins carries one $A_{\sigma}(\vec{q})$. Noticing the common factors for both orphan spins we can rewrite the effective action in terms of the undiluted correlators of the spin liquid so as to have effectively:
 \begin{align}
     S_{\text{int}}= \dfrac{D_1}{2(D_1^2-G_{12}^2)}(n_1^2+n_2^2)-\dfrac{G_{12}}{D_1^2-G_{12}^2}\vec{n}_1\cdot\vec{n}_2,
 \end{align}
 This action has now a constant term, since $n_1^2=n_2^2=1$ and an exchange term proportional to the charge charge correlator. We now use the assumption that $D_1\approx 1/\beta$  as we calculated in the previous section, valid for small magnetic field and low temperatures. We may approximate the final partition function as:
 \begin{align}
     Z_{\text{int}}= \int \mathcal{D} \vec{n}_1 \int \mathcal{D} \vec{n}_2 \ e^{ -\beta J_{\text{eff}}\ \vec{n}_1\cdot\vec{n}_2 }, \quad J_{\text{eff}}=-\dfrac{\beta}{3}\expval{\vec{\phi}_\gamma(\vec{r}_1)\cdot \vec{\phi}_\gamma(\vec{r}_2)}.
 \end{align}

This is indeed the result present in the first vacancy field theory paper \cite{sen_vacancy-induced_2012} . Let us then calculate the charge charge correlation function within the low-energy, long wavelength approximation. We can then use the previous equations for the inverse interaction matrix $h^{-1}(\vec{q})$ , to lowest order we obtain:
 \begin{align}
     &\dfrac{1}{3}\expval{\vec{\phi}_\gamma(\vec{r}_1)\cdot \vec{\phi}_\gamma(\vec{r}_2)}=\dfrac{1}{N}\sum_{\vec{q},\sigma,\sigma'}  e^{i\vec{q}\cdot \vec{R}_{21} } A_{\sigma}(\vec{q})h^{-1}_{\sigma\sigma'}(\vec{q})A_{\sigma'} (-\vec{q})\notag \\
    &\approx \dfrac{9}{32 (2\pi)^2}e^{i\vec{K}\cdot \vec{r}_{21}} \int^{\Lambda} \dd^2 \vec{k}\ \dfrac{(k_x+ik_y)^2(k_x-ik_y)^2}{\rho+g\beta  \vec{\abs{k}}^4}e^{i\vec{k}\cdot \vec{r}_{21}}+\dfrac{9}{32 (2\pi)^2}e^{i\vec{K}'\cdot \vec{r}_{21}} \int^{\Lambda} \dd^2 \vec{k}\ \dfrac{(k_x-ik_y)^2(k_x+ik_y)^2}{\rho+g\beta  \vec{\abs{k}}^4}e^{i\vec{k}\cdot \vec{r}_{21}}\\
    &\dfrac{1}{3} \expval{\vec{\phi}_\gamma(\vec{r}_1)\cdot \vec{\phi}_\gamma(\vec{r}_2)} =  \dfrac{9 T^{3/2}}{16 (2\pi)^2}\cos( \vec{K}\cdot \vec{r}_{21}) \int^{\Lambda/T^{1/4}} \dd^2 \vec{k}\ \dfrac{\vec{\abs{k}}^4}{\rho+g \vec{\abs{k}}^4}e^{i\vec{k}\cdot \vec{x}}.
 \end{align}
In the first equation we used the definition in terms of momentum space integrals. In the next line we expanded the $A_{\sigma}(\vec{q})$ as before and focused on the continuum limit of the sum.  Finally in the last line we combined everything into a single integral and got a cosine factor from considering both $\vec{K},\vec{K}'$ constributions. In the last tine we also defined again $\vec{x}=\vec{r}_{21}T^{1/4}$. We observe now an interesting scaling behaviour of the form:
\begin{align}
      \expval{\vec{\phi}_\gamma(\vec{r}_1)\cdot \vec{\phi}_\gamma(\vec{r}_2)}  \propto T^{3/2} F_2(\abs{\vec{r}_{1}-\vec{r}_{2}} T^{1/4}).
\end{align}

Similarly to the previous case we see that a different temperature prefactor appears, $T^{3/2}$ distinguishing our result from the usual Coulomb phase for the rank-1 $U(1)$ gauge theory which has $T^{1/2}$. Let us use now the Jacobi-Anger identity to rewrite the exponential in terms of Bessel functions:
\begin{align}
    \int_{0}^{\infty}\int_0^{2\pi}\text{k}  \dd \text{k}  \dd \theta  \ \ \frac{\text{k}^4}{\rho+g  \text{k}^4}e^{i\text{k} x\cos{(\theta-\phi)}}=\sum_{n=-\infty}^{\infty} \int_{0}^{\infty}\int_0^{2\pi}\text{k}  \dd \text{k}  \dd \theta  \ \ \frac{\text{k}^4 }{\rho+g  \text{k}^4} i^n J_{n}(\text{k} x)e^{in(\theta-\phi)}= 2\pi \int_{0}^{\infty}\dd \text{k}  \ \ \frac{\text{k}^5}{\rho+g  \text{k}^4} J_{0}(\text{k} x),
\end{align}
where in contrast to the previous section we have now only a contribution from the $n=0$ term in the sum , the reason for this is that the integrand has only $k^4$ dependence. We can again identify the last integral as the Hankel transform of order zero in $x$ which has the solution:
\begin{align}
    F_2(x)=-\frac{1024 G_{4,0}^{0,3}\left(\frac{256 g}{x^4 \rho }|
\begin{array}{c}
 -1,-1,-\frac{1}{2},-\frac{1}{2} \\
\end{array}
\right)}{\rho  x^6}.
\end{align}
To lowest order expanding near $x\ll1$ , which amounts to considering the separation between orphan clusters to be less than the correlation length, while still looking at large distances, we then obtain:
\begin{align}
      F_2(x)\approx-\frac{ \pi \sqrt{\rho } }{4g^{3/2}} +\frac{\rho  x^2 \left(\log \left(\frac{16 g}{\rho }\right)-4 (\log (x)+\gamma_1 -1)\right)}{16 g^2},
\end{align}
where the expansion naturally leads to the Euler-Mascheroni constant $\gamma_1$. Moreover we observe that a constant non decaying term is still present to the lowest finite temperatures.  If we assume the condition $x\ll1$ then the effective spin-spin correlation function between orphans is given by:
\begin{align}
    J_{eff}(\vec{r}_1,\vec{r}_2)= - \dfrac{\beta}{3}\expval{\vec{\phi}_\gamma(\vec{r}_1)\cdot \vec{\phi}_\gamma(\vec{r}_2)}\approx -\dfrac{18 \pi \sqrt{T} }{16  (2\pi)^2}\left(-\frac{64 \pi }{256 g^{3/2} \sqrt{\rho }}+\frac{\pi  \rho ^{3/2} x^4}{256g^{5/2}}\right)\cos( \vec{K}\cdot \vec{r}_{21}) = 0.56\sqrt{T}\cos( \vec{K}\cdot \vec{r}_{21}).
\end{align}
To lowest order there is again no decay of the spin spin exchange interaction, just as for the spin texture. In contrast the angular dependence present in the previous section is now missing. If one thinks of the correlator as the overlap of the pseudo-wave function $A_{\sigma}(\vec{q})$ , then the angular dependence cancels since it comes from a phase factor. The expectation value of the orphan spin-spin correlator is then given by:
\begin{align}
   \expval{\vec{S}(\vec{r}_1)\cdot \vec{S}(\vec{r}_2)}= \dfrac{1}{Z_{\text{int}}}\int \mathcal{D} \vec{n}_1 \int \mathcal{D} \vec{n}_2 \ \ \vec{n}_1\cdot \vec{n}_2 \ e^{ \beta K\ \vec{n}_1\cdot\vec{n}_2 }=  \dfrac{1}{\beta Z_{\text{int}}[K]} \dfrac{\partial }{\partial K }Z_{\text{int}[K]},
\end{align}
where we defined $K=-J_{eff}$ for convenience now the partition function can be calculated by just fixing one of the unit vectors  say $\vec{n}_1$ and integrating out $\vec{n}_2$. Integrating out the second spin then gives us the same partition function as a free spin in the field of $\vec{h}=K\vec{n}_1$ , which doesn't depend on the direction of $\vec{n}_1$ only the magnitude, which is $K$ so that we are left with the integrand being independent of $\vec{n}_1$ and so:
\begin{align}
    Z_{\text{int}}[K]= \int \mathcal{D} \vec{n}_1 Z_{\text{free}}[h=K] = 4\pi Z_{\text{free}}[h=K],
\end{align}
where we have used the independence on the direction of the partition function for a fixed $\vec{n}_1$ as mentioned before. The integral left is just the solid angle of the sphere which gives the factor $4\pi$.  The spin-spin correlation function is then of the same form as the magnetization calculated before in terms of the Langevin function $L(x)$ and is given by:
\begin{align}
     \expval{\vec{S}(\vec{r}_1)\cdot \vec{S}(\vec{r}_2)}= L(-\beta J_{\text{eff}})\approx L(- 0.56\cos( \vec{K}\cdot \vec{r}_{21})/\sqrt{T}).
\end{align}
We observe indeed no radial decay and an angular dependence coming only from the details of the lattice which selects the $\vec{K}$ vector to be the special momentum that characterizes the low energy behaviour of the system.

\end{document}